\documentclass[
 reprint,
 amsmath,amssymb,
 aps,
]{revtex4-1}

\usepackage{graphicx}
\usepackage{dcolumn}
\usepackage{bm}

\begin{document}

\title{Slowly damped quasinormal modes of the massive Dirac field in $d$-dimensional Tangherlini spacetime}

\author{Jose Luis Bl\'azquez-Salcedo}
\email[]{jose.blazquez.salcedo@uni-oldenburg.de}
\author{Christian Knoll}
\email[]{christian.knoll@uni-oldenburg.de}
\affiliation{Institut f\"ur Physik, Universit\"at Oldenburg, D-26111 Oldenburg, Germany}

\date{\today}

\begin{abstract}
We consider quasinormal modes of the massive Dirac field in the background of a Schwarzschild-Tangherlini black hole. Different dimensions of the spacetime are considered, from $d=4$ to $d=9$. The quasinormal modes are calculated using two independent methods: WKB and continued fraction. We obtain the spectrum of quasinormal modes for different values of the overtone number and angular quantum number. An analytical approximation of the spectrum valid in the case of large values of the angular quantum number and mass is calculated. Although we don't find unstable modes in the spectrum, we show that for large values of the mass, the quasinormal modes can become very slowly damped, giving rise to quasistationary perturbations.
\end{abstract}


\maketitle

\section{Introduction}

The properties of higher dimensional black holes have attracted much interest since several quantum gravity theories, such as string theory, brane world models, and the AdS/CFT correspondence, proposed the existence of more than four spacetime dimensions \cite{Emparan:2008eg,2012bhhd.book.....H}.
Of special interest in this context is the interaction of the black holes with several types of matter fields, in particular with test fields such as scalars and fermionic fields (Dirac spinors).

The interaction of such test fields in the background of a black hole has been widely investigated in the literature. In particular, the quasinormal mode decomposition of a field perturbation.
This analysis allows to calculate the resonant frequencies and damping times that govern the emission of radiation in a curved spacetime. In particular during the ringdown phase of a time-dependent field perturbation. The quasinormal mode analysis allows to test the mode stability of the solutions of the Einstein equations under small perturbations. In addition the spectrum has applications in the AdS/CFT correspondence, since the modes are related with the poles of the correlation functions. For a comprehensive review see for example \cite{Konoplya:2011qq,Ferrari:2007dd,Berti:2009kk}.

Focusing on the fermionic fields, the analysis of the Dirac equation benefits from the fact that it is known to be separable  into radial and angular parts in a number of geometries describing rotating black holes. The separability has been related to hidden symmetries of the metric background \cite{Frolov:2017kze,PhysRevD.19.1093}, and introduces an angular operator with its corresponding quantum number. This is well known for the Kerr black hole \cite{10.2307/79011,1984RSPSA.391...27C}, and it also holds in the presence of charge \cite{Mukhopadhyay:2000ss} and cosmological constant \cite{Belgiorno:2008hk}. In higher dimensions, 
the separability has been studied for the 5D Myers-Perry black hole \cite{Wu:2008df} and in the more general case of the higher dimensional Kerr-NUT-dS black hole
\cite{Oota:2007vx}.

Regarding hidden symmetries, the Killing-Yano tensors for the most general charged rotating geometries was constructed in \cite{Chervonyi:2015ima}. The separation of the Maxwell equations in the Myers-Perry geometry using Killing-Yano tensors was recently achieved in \cite{Lunin:2017drx}.

Naively one may be tempted to think that 
a Dirac spinor could be excited
to form a stable configuration around a black hole.
Such a configuration would mimic an atom, with the event horizon surrounded by a stationary fermionic field \cite{0143-0807-28-3-007}. 
However it has been shown that, 
under some generic assumptions, 
there do not exist stable solutions to the Einstein-Dirac equations with fermionic hair, even when supplementing the system with other fields as well \cite{Finster:1999mn,Finster:1998ak,Finster:1999ry,Finster:2002vk}. The spinor field either falls into the black hole or vanishes at infinity (being radiated away from the horizon). To further cement this point, there are various proofs for the nonexistence of purely real frequencies in the spectrum of quasinormal modes of the Dirac field. For the Schwarzschild spacetime see \cite{Batic:2017qcr,Lasenby:2002mc} and for the five dimensional Myers-Perry spacetime see \cite{Daude:2012wq}. However there are known examples of exotic stable configurations, which have been proposed as dark matter contributors \cite{Dokuchaev:2013kra,Dokuchaev:2014vda}. {Let us note here that, recently, Dirac stars (the fermionic equivalent of boson and Proca stars, with no horizon) have been constructed, using a pair of fermionic fields instead of a single field \cite{Herdeiro:2017fhv}.}

The nonexistence of stable solutions to the Einstein-Dirac equations with fermionic hair and the lack of superradiance for spinors \cite{Brito:2015oca,Maeda:1976tm,Iyer:1978du} indicates that, in terms of the quasinormal mode analysis, the perturbations in the Dirac field will decay with time.
The massless quasinormal mode spectrum of Dirac spinors in the geometry of the Schwarzschild black hole was investigated in \cite{Jing:2005dt} using the continued fraction and Hill-determinant methods. They found that the fundamental quasinormal modes become evenly spaced for large angular quantum number. The quasinormal modes also become evenly spaced 
as the overtone number is increased.
It was also shown that
the angular quantum number affected the real part of the frequencies but had almost no influence on the imaginary part.
The massive modes were investigated in \cite{Cho:2003qe} using a WKB approach, although only small mass values of the field were studied. 
It was observed that when the mass is increased, the real part of the quasinormal frequencies increases as well, but interestingly the absolute value of the imaginary part decreases. This indicates that massive Dirac fields decay slower in time than in the massless case. 

The massless modes in the higher dimensional Schwarzschild/Tangherlini spacetime were investigated in \cite{Cho:2007zi} using a WKB approach. 
In this study it is shown that for higher dimensions the damping time increases. Using the P\"oschl-Teller potential approximation the modes of the massless field were also studied in the Reissner-Nordstr\"om de-Sitter black hole \cite{Jing:2003wq} and in the Schwarzschild-de Sitter black hole \cite{Zhidenko:2003wq} (here also the WKB approximation up to sixth order was used). In both cases an increase in the cosmological constant led to a decrease in the absolute value of the imaginary part of the quasinormal frequencies, making the modes more slowly damped. For Schwarzschild-de Sitter it was mentioned that an increase in the cosmological constant also led to a decrease of the real part of the frequencies. For the Reissner-Nordstr\"om-de-Sitter black hole the absolute value of the frequency decreased when increasing the angular quantum number, but became larger when increasing the charge of the black hole and the overtone number. An analytical investigation in the asymptotic spectrum (meaning high overtone number) of a Dirac field in the geometry of the Schwarzschild-AdS spacetime with numerical checking of the results was done in \cite{Arnold:2013gka}.
The quasinormal modes of Weyl spinors in the BTZ black hole were calculated in \cite{Cardoso:2001hn}.

Investigation of the quasinormal modes for the massive field in the geometry of the Kerr black hole was carried out in \cite{Dolan:2015eua}. One of the interesting results is that in the rapidly rotating case the decay rate of low frequency co-rotating quasinormal modes are suppressed in the (bosonic) superradiant regime.
The scattering of massive Dirac fields in the Schwarzschild black hole was investigated in \cite{0264-9381-15-10-018}. Analytical expressions for the phase shift in the scattering of massive fermions can be found for the Schwarzschild black hole in \cite{Cotaescu:2014jca} and for the Reissner-Nordstr\"om black hole in \cite{Cotaescu:2016aty}.
 Recently, analytical solutions were obtained in \cite{Sporea:2015wsa} describing quasinormal modes in the near-horizon regime. 

As we mentioned above, 
it has been generally noted that increasing the mass of a field leads to longer lived modes. In particular, for scalar fields this was noted in \cite{Konoplya:2004wg,0264-9381-9-4-012,Zhidenko:2006rs} and for vector fields in \cite{Konoplya:2005hr}. 
In the case of vector fields, some quasinormal modes show the curious behaviour of decreasing the frequency as the mass increases, eventually becoming a pulse. For large enough values of the vector field mass the modes can cease existing. 

In this paper we will investigate the quasinormal modes of the Dirac field in the geometry of the $d$-dimensional Tangherlini spacetime. The focus will be on the behaviour of the field for large masses. We will see that indeed, the field follows this general behaviour, with longer lived modes for larger values of the mass. 

The structure of the paper will be as follows. In Section \ref{S1}, first we will introduce the conventions we use in the paper, and then we will derive the differential equations governing the radial part of the field, with a study of the asymptotic behaviour of the perturbation. In order to generate the quasinormal modes, we will use two independent methods: the method of continued fractions with the Nollert improvement \cite{Leaver:1985ax, Nollert:1993zz} and the WKB method up to third order \cite{Iyer:1986np}, presenting in both cases all the necessary equations in Section \ref{S2}. 
In Section \ref{S3} we present the results, where we start with the analysis for large angular quantum number. Here we combine the numerical methods with analytical results obtained in the limit of large angular quantum number and large fermionic mass. 
We also present results for $l=0$, studying the fundamental state and the first excitation. In Section \ref{S4} we finish with some conclusions and an outlook.
 
\section{Dirac equation in Tangherlini spacetime}\label{S1}

Let us begin with a short note on conventions. Our sign convention for the metric of special relativity is $\eta \stackrel{*}{=} \mathrm{diag}[1,-1, \dots, -1]$. We will use the Einstein summation convention, always summing over the whole possible range for the indices when not otherwise stated. We will use greek letters for coordinate components of tensors and latin letters for
components in the orthonormal frame,
 for example $\mathbf{g} = g_{\mu \nu} \mathbf d x^\mu \otimes \mathbf d x^\nu = \eta_{a b} \bm{\omega}^a \otimes \bm{\omega}^b$. To distinguish between components in the coordinates and in the orthonormal frame we will give components in the frame a hat, 
so $v_i$ is the $i$-th component of $\mathbf{v}$ in the coordinates and $v_{\hat i}$ is the $i$-th component of $\mathbf{v}$ in the orthonormal frame.

The metric of the $d$-dimensional Schwarzschild-Tangherlini spacetime is
\begin{eqnarray}
\mathrm d s^2 = f(r) \, \mathrm d t^2 - \frac{1}{f(r)} \, \mathrm d r^2 - r^2 \, \mathrm d \Omega^2_{d-2} \, ,
\end{eqnarray}
with $f(r) = 1-(\mu/r)^{d-3}$, $\mu$ being related to the mass of the black hole and $\mathrm d \Omega^2_{d-2}$ being the line element of the $d-2$ dimensional sphere. We use the vielbein
\begin{eqnarray}
\bm{\omega}^{\hat 0} = \sqrt{f(r)} \, \mathbf d t \, , \; \bm{\omega}^{\hat 1} = \frac{1}{\sqrt{f(r)}} \, \mathbf d r \, , \; \bm{\omega}^{\widehat{i+1}} = r \, \bm{\omega}^{\hat i}_{d-2} \, ,
\end{eqnarray}
with $\bm{\omega}^{\hat i}_{d-2}$ being a vielbein of the $d-2$ dimensional sphere and $i$ ranging from 1 to $d-2$. This allows us to write the Dirac equation
\begin{eqnarray}
\mathcal{D} \Psi = && \left( \frac{\mathrm{i}}{\sqrt{f}} \gamma^{\hat 0} \partial_t + \mathrm{i} \sqrt{f} \gamma^{\hat 1} \left[ \partial_r + \frac{\mathrm d}{\mathrm d r} \ln r^{(d/2) - 1} f(r)^{1/4} \right] \right. \nonumber \\
&& \left. + \frac{\mathrm i}{r} \gamma^{\hat 0} \gamma^{\hat 1} \mathcal{K}_{d-2} - m \mathbb{E} \right) \Psi = 0 \, ,
\end{eqnarray}
with $m$ being the mass of the Dirac field, $\mathcal{K}_{d-2}$ being the angular operator and $\mathbb{E}$ being the unit operator. 

It is true that $[\mathcal{D},\mathcal{K}_{d-2}]=0$, so that one can split the spinor as follows
\begin{eqnarray}
\Psi = \frac{r}{r^{d/2} f(r)^{1/4}} \, \mathrm{e}^{- \mathrm{i} \omega t} \, \phi(r) \otimes \Theta_\kappa \, ,
\label{Psi_exp}
\end{eqnarray}
with \cite{CAMPORESI19961}
\begin{eqnarray}
\mathcal{K}_{d-2} \Theta_\kappa = \kappa  \Theta_\kappa = \pm \left(l + \frac{d-2}{2} \right) \Theta_\kappa \, ,
\label{kappa_def}
\end{eqnarray}
and $l \in \mathbb N_0$ being the angular quantum number.
It is worth mentioning that, for a given value of $l$, $\kappa$ can be positive or negative, and we will see that in the case of massive fields, each sign results in a branch of modes that possess different properties.

In addition, from equation (\ref{Psi_exp}) we can see that we are focusing on a mode decomposition of the time dependent perturbation by introducing the eigenfrequency $\omega$. With this Ansatz the resulting differential equation for the radial part $\phi$ is
\begin{eqnarray}
 \left( \frac{\omega}{\sqrt{f}} \gamma^{\hat 0} + \mathrm{i} \sqrt{f} \gamma^{\hat 1}  \frac{\mathrm d}{\mathrm d r}  + \frac{\mathrm i \kappa}{r} \gamma^{\hat 0} \gamma^{\hat 1}  - m \mathbb{E} \right) \phi = 0 \, .
 \label{radial_eq_1}
\end{eqnarray}

It is convenient to work with normalised quantities such as
\begin{eqnarray}
x = r / \mu \, , \; \Omega = \mu \omega \, , \; \eta = \mu m \, ,
\end{eqnarray}
which simplifies equation (\ref{radial_eq_1}) to
\begin{eqnarray}
 \left( \frac{\Omega}{\sqrt{f}} \gamma^{\hat 0} + \mathrm{i} \sqrt{f} \gamma^{\hat 1}  \frac{\mathrm d}{\mathrm d x}  + \frac{\mathrm i \kappa}{x} \gamma^{\hat 0} \gamma^{\hat 1}  - \eta \mathbb{E} \right) \phi = 0 \, .
 \label{radial_eq_2}
\end{eqnarray}
In addition, it can be convenient to change variables to the tortoise coordinate $z$, defined by the relation
\begin{eqnarray}
\frac{\mathrm d}{\mathrm d z} = f \frac{\mathrm d}{\mathrm d x},
\end{eqnarray}
which transforms equation (\ref{radial_eq_2}) to
\begin{eqnarray}
 \left( \Omega \gamma^{\hat 0} + \mathrm{i}  \gamma^{\hat 1}  \frac{\mathrm d}{\mathrm d z}  + \frac{\mathrm i \kappa \sqrt{f}}{x} \gamma^{\hat 0} \gamma^{\hat 1} - \eta \sqrt{f} \, \mathbb{E} \right) \phi = 0 \, . \label{eqn:diff.eq-tortoise}
\end{eqnarray}

In order to study the quasinormal modes of this system, we need to set 
a number of physically relevant boundary conditions at the horizon and at spatial infinity.
It is convenient to express these conditions in terms of the
probability current $\mathbf j$, which has to be conserved, $\mathrm d \ast \mathbf j = 0$. 
Let us choose as
four dimensional volume $\cal{V}$ the spatial hypersurface  $V(t)$ orthogonal to the Killing vector $\partial_t$ outside the horizon  $\cal{H}$ of the black hole, an $\epsilon$-distance away from the horizon, translated for a time $\Delta t$ from an initial time $t_0$.
Then we can express the conservation law as
\begin{eqnarray}
0 = \int_{\cal{V}}  \mathrm d \ast \mathbf j  = \int_{\partial \cal{V}} \ast \mathbf j =&& \underbrace{\left( \int_{V(t=t_0 + \Delta t)} - \int_{V(t=t_0)} \right) \ast \mathbf j}_{<0} \nonumber \\
&&+ \int\limits_{S_{d-2}^\infty \times \Delta t} \ast \mathbf{j} - \int\limits_{{\cal{H}_\epsilon} \times \Delta t} \ast \mathbf{j} \, ,
\end{eqnarray}
where $S_{d-2}^\infty$ is a $d-2$-sphere at spatial infinity and $\mathcal{H}_\epsilon$ is a $d-2$-sphere at $r=\mu + \epsilon$. We know that the first summand is smaller than zero, because we will have a decaying field. Thus the second and third terms together must be greater than zero. To achieve this, one possibility is to require the following conditions to the probability current:
\begin{eqnarray}
\int\limits_{{\cal{H}_\epsilon} \times \Delta t} \ast \mathbf{j} \, &&=  \int\limits_{\Delta t} \mathrm d t \,  \left. j^r \right|_{r=\mu+\epsilon} \int\limits_{\cal{H}_\epsilon} \mathrm{d} \Sigma_r 
<
 0 \, , \nonumber \\
\int\limits_{S_{d-2}^\infty \times \Delta t} \ast \mathbf{j} \, &&=  \int\limits_{\Delta t} \mathrm d t \,  \left. j^r \right|_{r \rightarrow \infty} \int\limits_{S_{d-2}^\infty} \mathrm{d} \Sigma_r 
>
 0 \, .
\label{bc_flux}
\end{eqnarray}
The integration surfaces are intrinsic geometric objects being generated by the orbits of  Killing vectors.
The requirements to the 
current at the boundaries (\ref{bc_flux})
imply that the field flows into the black hole at the horizon 
($r \rightarrow \mu \Rightarrow x \rightarrow 1 \Rightarrow z \rightarrow - \infty$)
, meaning that $j^r < 0$ there (we have taken the limit $\epsilon \rightarrow 0$ here). At spatial infinity ($r \rightarrow \infty \Rightarrow x \rightarrow \infty \Rightarrow z \rightarrow \infty$) the field should flow outward, so $j^r > 0$ there.

Let us now choose the following representation for the Clifford algebra and the spinor
\begin{eqnarray}
\gamma^{\hat 0} = \left[ \begin{array}{cc} 0 & 1 \\ 1 & 0 \end{array} \right] \, , \; \gamma^{\hat 1} = \left[ \begin{array}{cc} 0 & 1 \\ -1 & 0 \end{array} \right] \, , \; \phi = \left[ \begin{array}{c} \phi_1 \\ \phi_2 \end{array} \right], \label{eqn:Clifford-representation}
\end{eqnarray}
which is appropriate in order to simplify several expressions, in particular later in section III. For instance, with this choice the radial probablity current 
$j^{\hat 1} = j^{\hat r} \propto j^r$ is 
just 
proportional to $| \phi_2 |^2 - | \phi_1 |^2$. 
This allow us to rewrite the requirements from
equation (\ref{bc_flux})
as a 
set of boundary conditions for the radial part of the spinor, which constrains the behaviour of the leading terms at the boundaries.

At the horizon, from equations (\ref{bc_flux}) and (\ref{eqn:Clifford-representation}) we obtain that $| \phi_2 |^2 - | \phi_1 |^2<0$, which implies that the radial part of the spinor has to behave like
\begin{eqnarray}
\left[ \begin{array}{c} \phi_1 \\ \phi_2 \end{array} \right] && \approx \left[ \begin{array}{c} 1 \\ \frac{2 \mathrm{i} \sqrt{d -3}}{4 \mathrm{i} \Omega - d + 3} \, \mathrm{e}^{\frac{d-3}{2}  z} \end{array} \right]  \mathrm{e}^{- \mathrm{i} \Omega z} \nonumber \\ 
&&\approx \left[ \begin{array}{c} 1 \\ \frac{2 \mathrm{i} \sqrt{d -3}}{4 \mathrm{i} \Omega - d + 3} \, \sqrt{x-1} \end{array} \right] (x-1)^{- \frac{\mathrm{i} \Omega}{d - 3}} \, .
\label{bc_horizon}
\end{eqnarray}

Similarly, at
infinity, equations (\ref{bc_flux}) and (\ref{eqn:Clifford-representation}) implies that $| \phi_2 |^2 - | \phi_1 |^2>0$, which results in the following behaviour of the radial part of the spinor
\begin{eqnarray}
\left[ \begin{array}{c} \phi_1 \\ \phi_2 \end{array} \right] && \approx \left[ \begin{array}{c} \Omega \pm \chi \\ \eta \end{array} \right]  z^{\mp \alpha_z} \, \mathrm{e}^{\mp \mathrm{i} \chi z} \nonumber \\
&& \approx  \left[ \begin{array}{c} \Omega \pm \chi \\ \eta \end{array} \right] x^{\mp \alpha_x} \, \mathrm{e}^{\mp \mathrm{i} \chi x} \, ,
\label{bc_infinity}
\end{eqnarray}
with $\chi = \sqrt{\Omega^2 - \eta^2}$, the upper sign for $\Re (\Omega)<0$, the lower sign for $\Re (\Omega) > 0$ and the constants $\alpha_z$ and $\alpha_x$ are defined such that
\begin{eqnarray}
\alpha_z &&= \begin{cases} \frac{\mathrm i}{2} \frac{\eta^2}{\chi} \; &\text{, for} \; d=4 \\ 0 &\text{, otherwise} \end{cases} \, , \nonumber \\
\alpha_x &&= \begin{cases} \alpha_z + \mathrm i \chi = \frac{\mathrm i}{2} \frac{2 \Omega^2 - \eta^2}{\chi}  &\text{, for} \; d=4  \\ 0 &\text{, otherwise} \end{cases} \, . 
\end{eqnarray}

Note that $d=4,5$ present a distinct behaviour asymptotically.
When the representation given in equation (\ref{eqn:Clifford-representation}) is used with the differential equation (\ref{eqn:diff.eq-tortoise}), it results in the following second order differential equation for the functions $\phi_{1,2}$
\begin{eqnarray}
(\mathrm{i} \kappa &&\mp x \eta) x^2 \frac{\mathrm d^2}{\mathrm d z^2} \phi_{1,2} + x F_\mp \frac{\mathrm d}{\mathrm d z} \phi_{1,2} + \left\{ (\mathrm i \kappa \mp x \eta) x^2 \Omega^2 \right. \nonumber \\
&& \left. \pm \mathrm i \Omega x F_\mp  - (\mathrm i \kappa \mp \eta x) (\kappa^2 + \eta^2 x^2) f \right\} \phi_{1,2} = 0  \, , \label{eqn:diff.second.order}
\end{eqnarray}
with
\begin{eqnarray}
F_\mp (z) = \mathrm i \kappa  f - \frac{x (\mathrm i \kappa \mp \eta x)}{2 f} \frac{\mathrm d f}{\mathrm d z} \, ,
\end{eqnarray}
the upper sign for $\phi_1$ and the lower sign for $\phi_2$. {Note that since we are working with another representation, this differential equation is different from the one obtained in \cite{Cho:2003qe}.}

\section{Numerical methods}\label{S2}

Because of the lack of analytical solutions to the equation (\ref{eqn:diff.second.order}) subject to the boundary conditions (\ref{bc_horizon}) and (\ref{bc_infinity}), we need to employ numerical methods in order to obtain the spectrum of quasinormal modes of the massive Dirac field.

We will employ two independent techniques in our calculations: the continued fraction method with the Nollert improvement, and a third order WKB method. 

\subsection{Continued fraction method}

From the asymptotic behaviour of the spinor that we have obtained in equation (\ref{bc_infinity}), we
can factorize the 
behaviour of the Ansatz functions at infinity
\begin{eqnarray}
\left[ \begin{array}{c} \phi_1 \\ \phi_2 \end{array} \right] =  x^{\alpha_x} (x-1)^{-\frac{\mathrm{i} \Omega}{d-3}} \mathrm{e}^{ \mathrm{i} \chi x}  \left[ \begin{array}{c} \psi_1 \\ \sqrt{x-1} \, \psi_2 \end{array} \right] \, ,
\end{eqnarray}
with $\psi_1$ and $\psi_2$ unknown functions of $x$, and where we have assumed the $\Re(\Omega) > 0$ behaviour in the various exponents (i.e. the lower sign in equation (\ref{bc_infinity})). In addition, it is convenient to
change variables to the compactified coordinate
$y := 1 - \frac{1}{x}$
with $y \in [0,1] $.
Then the
second order differential equation (\ref{eqn:diff.second.order}) 
gives as a result the following system of equations for $\psi_1$ and $\psi_2$:

\

\

\begin{eqnarray}
&&K_\mp (1-y)^4 f^2 \frac{\mathrm d^2}{\mathrm d y^2} \psi_{1,2} \nonumber \\
&&+ \Bigg\{ G K_\mp \mp \mathrm i \kappa (1-y)^2 f \Bigg\} (1-y)^2 f \frac{\mathrm d}{\mathrm d y} \psi_{1,2} \nonumber \\
&&+ \Bigg\{ \left( C f^2 + \Omega^2 \mp \frac{\mathrm i \Omega}{2} (1-y)^2 \frac{\mathrm d f}{\mathrm dy} - f K_- K_+ \right. \nonumber \\
&& \;\;\;\;\;\;\;\;\; \left.  + \frac{1}{2} B (1-y)^2 f \frac{\mathrm d f}{\mathrm d y} \right) K_\mp \nonumber \\
&& \;\;\;\;\;\;+ \left( \Omega \mp \mathrm i B f \right) \kappa (1-y)^2 f \Bigg\} \psi_{1,2} = 0 \, ,  \label{eqn:CF}
\end{eqnarray}
with the upper sign for $\psi_1$, the lower sign for $\psi_2$ and for simplicity we have defined the following functions

\begin{eqnarray}
K_\pm &&= \eta \pm \mathrm i \kappa (1-y) \, , \, A = (x-1)^{\tilde \alpha} x^{\alpha_x} \mathrm e^{\mathrm i \chi x} \, , \nonumber \\
B &&= \frac{\mathrm d}{\mathrm d x} \ln A = - \frac{\mathrm i \Omega}{d - 3} \frac{1-y}{y} + \alpha_x (1-y) + \mathrm i \chi \, , \nonumber \\
C &&= \frac{1}{A} \frac{\mathrm d^2}{\mathrm d x^2} A \nonumber \\
&&= \left( \left[ \frac{\tilde{\alpha}}{y} + \alpha_x \right] (1-y) + \mathrm i \chi \right)^2 - \left[ \frac{\tilde{\alpha}}{y^2} + \alpha_x \right] (1-y)^2 \, , \nonumber \\
G &&= 2f(B+y-1)+\frac{1}{2}(1-y)^2\frac{df}{dy}
\nonumber \\
\tilde{\alpha} &&= \begin{cases} - \frac{\mathrm i \Omega}{d - 3} &\text{, for} \; \psi_1 \\ - \frac{\mathrm i \Omega}{d - 3} + \frac{1}{2}  &\text{, for} \; \psi_2 \end{cases} \, .
\end{eqnarray}

Note that the zeros of the coefficient of $\mathrm d^2 \psi_{1,2} / \mathrm d y^2$ for $4 \le d \le 9$ are either at $y=0$ or outside the unit circle $|y|<1$ in the complex $y$-plane. However, for $d>9$ the zeros of the function $f(y) = 1 - (1-y)^{d-3}$ will lie inside the unit circle. Thus one can expand the functions $\psi_{1,2}$ in a power series in $y$, convergent on the whole range $y \in [0,1]$ of interest on the real axis only for $4 \le d \le 9$. For $d \ge 10$ one has to analytically continue the functions through midpoints \cite{Rostworowski:2006bp}. We will not do this here, hence in the following we will restrict to the cases $4 \le d \le 9$. All coefficient functions of these differential equations are just polynomials in $y$. Thus the resulting recurrence relations for the coefficients of the expansion will be of finite order, namely $2 d - 3$.

We will use the continued fraction method to determine the complex frequencies $\Omega$ which lead to physical solutions of the system (\ref{eqn:CF}) \cite{Leaver:1985ax}. The method with the Nollert improvement is described in \cite{Nollert:1993zz}. Given a recurrence relation of order $N$ for the coefficients $f_{n}$
\begin{eqnarray}
\sum\limits_{k=0}^{\min \{ N, n\}} a^{(N)}_k (n) \, f_{n-k} = 0 \, ,
\end{eqnarray}
one can calculate the coefficients of the recurrence relation of order $N-1$. These coefficients for $n < N, 0 \le k \le n$ are given by
$a^{(N-1)}_k (n) = a^{(N)}_k (n)$  \, ,
and for $n \ge N$ by
\begin{widetext}
\begin{equation}
\left[ \begin{array}{c} a^{(N-1)}_0 (n) \\ a^{(N-1)}_1 (n) \\ a^{(N-1)}_2 (n) \\ \vdots \\ a^{(N-1)}_{N-2} (n) \\ a^{(N-1)}_{N-1} (n) \end{array} \right] = \left[ \begin{array}{cccccc} 0 & 0 & 0 & \cdots & 0 & a^{(N)}_0 (n) \\ - a^{(N)}_N (n) & 0 & 0 & \cdots & 0 & a^{(N)}_1 (n) \\ 0 & -a^{(N)}_N (n) & 0 & \cdots & 0 & a^{(N)}_2 (n) \\ \vdots & & \ddots & & \vdots & \vdots \\ 0 & \cdots & 0 & -a^{(N)}_N (n) & 0 & a^{(N)}_{N-2} (n) \\ 0 & \cdots & 0 & 0 & -a^{(N)}_N (n) & a^{(N)}_{N-1} (n) \end{array} \right] \, \left[ \begin{array}{c} a^{(N-1)}_0 (n-1) \\ a^{(N-1)}_1 (n-1) \\ a^{(N-1)}_2 (n-1) \\ \vdots \\ a^{(N-1)}_{N-2} (n-1) \\ a^{(N-1)}_{N-1} (n-1) \end{array} \right] \,.
\end{equation}
\end{widetext} 
To avoid large numbers in the numerics one can normalise after each step by dividing for example with $a_{N-1}^{(N-1)}(n-1)$, provided that  $a_{N-1}^{(N-1)}(n-1) \neq 0$.

Thus it is possible to reduce the recurrence relation to a relation of order two
\begin{eqnarray}
a^{(2)}_0 (1) \, f_1 + a^{(2)}_1 (1) \, f_0 &&= 0  \, , \nonumber \\
a^{(2)}_0 (n) \, f_n + a^{(2)}_1 (n) \, f_{n-1} + a^{(2)}_2 (n) \, f_{n-2} &&= 0  \, .
\end{eqnarray}
with $n \ge 2$. Re-expressing this with $\Delta_n := f_n / f_{n-1}$ gives
\begin{equation} \label{eqn:Delta.1}
a^{(2)}_0 (1) \, \Delta_1 + a^{(2)}_1 (1) = 0  \, ,
\end{equation}
\begin{equation}\label{eqn:Delta.n}
\Delta_{n-1} = \frac{- a^{(2)}_2(n)}{a^{(2)}_1 (n) + a^{(2)}_0 (n) \, \Delta_n} \, ,
\end{equation}
with $n \ge 2$. Using equation (\ref{eqn:Delta.n}) in equation (\ref{eqn:Delta.1}) results in the following continued fraction equation 
\begin{eqnarray}
a^{(2)}_1 (1) \,  - \, \frac{a^{(2)}_0(1) \, a^{(2)}_2(2)}{a^{(2)}_2(2) -  \frac{a^{(2)}_0 (2) \, a^{(2)}_2 (3)}{a^{(2)}_1 (3) - 
\raisebox{-2.5pt}{$\ddots$} 
\frac{a^{(2)}_0 (n-1) \, a^{(2)}_2 (n)}{a^{(2)}_1(n) -  
\raisebox{-3.0pt}{$\ddots$} 
}}} = 0 \, . \label{eqn:contfract}
\end{eqnarray}

The coefficients $a^{(2)}_k(n)$ will depend on $\Omega$. Thus one can approximate the above continued fraction up to a certain depth and try to minimize the above difference by changing $\Omega$. Let the depth of the approximation be $K-1$. Then the last fraction will be given by
\begin{eqnarray}
\raisebox{-2.5pt}{$\ddots$} 
\frac{a^{(2)}_0 (K-1) \, a^{(2)}_2 (K)}{a^{(2)}_1 (K) + a_0^{(2)} (K) \, \Delta _K} \, .
\end{eqnarray}
Instead of approximating $\Delta_K$ by zero, the Nollert improvement uses the original recurrence relation to give an asymptotic approximation of $\Delta_K$ in $K \gg 1$.

The coefficients $a^{(2d-1)}_k (K)$ and $\Delta_K$ are expanded in powers of $1/\sqrt{K}$ and $K-k \approx K$ for $0 \le k \le 2d-1 \ll K$ is used. Let the expansion of $\Delta_K$ be
\begin{eqnarray}
\Delta_K = C_0 + \frac{C_1}{\sqrt{K}} + \mathcal{O} \left( \frac{1}{K} \right).
\end{eqnarray}
Following Nollert \cite{Nollert:1993zz} we choose $C_0 = -1$ and $\Re ( C_1 ) > 0$. The other coefficients are then uniquely determined by the resulting equations.

By using this approach with equation (\ref{eqn:CF}), we obtain two continued fraction relations. 
Unless stated otherwise, to generate the quasinormal mode frequencies $\Omega$, we 
search with both equations separately, due to 
the lack of an obvious supersymmetry
between the two equations when the fermionic mass is not zero \cite{Cho:2007zi}.

The method is implemented in Maple. The initial mass value is $\eta = 10^{-6}$. The initial guess for the initial mass value  in the complex $\Omega$-plane is chosen close to the large $\kappa$ analytic approximation for massless modes from \cite{Cho:2007zi}. The ratio of the lefthand side of equation (\ref{eqn:contfract}) at a resonance to the surrounding area in the complex $\Omega$-plane is $\mathcal O(10^{-4})$. 

\subsection{WKB Method}

In addition to the continued fraction approach, we will use a third order WKB method.
In this case it is convenient to factorize
the spinor as
\begin{eqnarray}
\phi_{1,2} = \mathrm{e}^{g_\mp} \, \nu_{1,2} \, ,
\end{eqnarray}
with
\begin{eqnarray}
\frac{\mathrm d}{\mathrm d z} g_\mp = - \frac{F_\mp}{2 x (\mathrm i \kappa \mp x \eta)} \, ,
\end{eqnarray}
the upper sign for $\phi_1$ and the lower sign for $\phi_2$. Equation (\ref{eqn:diff.second.order}) is then reduced to the second order differential equation
\begin{eqnarray}
\frac{\mathrm d^2}{d z^2} \nu_{1,2} + \left( \Omega^2 - V_{\text{eff}, \mp} \right) \nu_{1,2} = 0 \, ,
\label{eq_WKB}
\end{eqnarray}
with the effective potential
\begin{eqnarray}
V_{\text{eff},\mp} = &&\mp \frac{\mathrm i \Omega}{x ( \mathrm i \kappa \mp x \eta)} \, F_\mp + \frac{1}{2} \left(\frac{F_\mp}{2 x (\mathrm i \kappa \mp x \eta)}  \right)^2 \nonumber \\
&&+ \frac{1}{2} \frac{\mathrm d}{\mathrm d z} \left( \frac{F_\mp}{2 x (\mathrm i \kappa \mp x \eta)}  \right) + \frac{\kappa^2 + \eta^2 x^2}{x^2} \, f \label{eqn:completeeffpot}
\end{eqnarray}
where again the upper sign is for $\nu_1$ and the lower sign is for $\nu_2$. This second order equation is in the gestalt of a Schr\"odinger-like equation. The boundary conditions are outgoing waves at $z \rightarrow \pm \infty$. This problem was solved semi-analytically in \cite{Iyer:1986np} with a WKB approach up to third order. The equation for the complex frequencies $\Omega$ is \cite{Iyer:1986nq,0264-9381-9-4-012}
\begin{eqnarray} 
\Omega^2 = [V_0 + ({-2V_0^{(2)}})^{1/2}\Lambda ] - \mathrm i \lambda (-2 V_0^{(2)} )^{1/2} (1 + \Sigma) \, , \label{eqn:WKBeqn}
\end{eqnarray}
where we have defined the following functions:
\begin{eqnarray}
\Lambda = && \frac{1}{({-2V_0^{(2)}})^{1/2}}\left\{\frac{1}{8} \left( \frac{V_0^{(4)}}{V_0^{(2)}} \right) \left( \frac{1}{4} + \lambda^2 \right) \right.  \nonumber \\
&& \left. - \frac{1}{288} \left( \frac{V_0^{(3)}}{V_0^{(2)}} \right)^2 (7 + 60 \lambda^2)\right\}  \, , \nonumber \\
\Sigma = && \frac{-1}{2 V_0^{(2)}} \left\{ \frac{5}{6912} \left( \frac{V_0^{(3)}}{V_0^{(2)}} \right)^4 (77 + 188 \lambda^2) \right. \nonumber \\
&& - \frac{1}{384} \left( \frac{V_0^{(3) \, 2} V_0^{(4)}}{V_0^{(2) \, 3}} \right) (51 + 100 \lambda^2) \nonumber \\
&& + \frac{1}{2304} \left( \frac{V_0^{(4)}}{V_0^{(2)}} \right)^2 (67 + 68 \lambda^2 ) \nonumber \\
&& + \frac{1}{288} \left( \frac{V_0^{(3)} V_0^{(5)}}{V_0^{(2) \, 2}} \right) (19 + 28 \lambda^2) \nonumber \\
&& \left. - \frac{1}{288} \left( \frac{V_0^{(6)}}{V_0^{(2)}} \right) (5 + 4 \lambda^2) \right\} \, ,
\end{eqnarray}
$\lambda = n + 1/2$ and $V_0^{(k)}$ being the $k$'th derivative of $V_{\text{eff}, \mp}$ evaluated at the point where the extremum $\mathrm d / \mathrm d z \, V_{\text{eff}, \mp} = 0$ is found. It should be noted, that the WKB method had been improved up to 6th order in \cite{Konoplya:2003ii} and recently modified and improved up to 13th order in \cite{Matyjasek:2017psv}.

The potential $V_{\text{eff}, \mp}$ is complex, thus one cannot expect to find a point with  $\mathrm d / \mathrm d z \, V_{\text{eff}, \mp} = 0$ on the real $z$-axis. We will thus 
make an analytical continuation of
$z$ to the complex plane. We will also search using $x$ instead of $z$, because it is not possible to give an analytically closed expression for $x=x(z)$ for all dimensions $d$. 

The method is implemented in Maple.  For each given set of parameters we first numerically calculate the location of the extremum $x_0$ in the complex $x$-plane. From the set of values $x_0$ for which $\Re(x_0) > 1$ we choose the one with the smallest $|\Im(x_0)|$. After the extremum is found, the left- and righthand sides of equation (\ref{eqn:WKBeqn}) are evaluated and the absolute value of the difference of these is calculated.
The search in the complex $\Omega$-plane aims to minimize this difference. The initial mass value is $\eta = 0$. The initial guess for the initial mass value in the complex $\Omega$-plane is chosen according to the large $\kappa$ analytic approximation for massless modes from \cite{Cho:2007zi}. The ratio of the difference of the left- and righthandside of equation (\ref{eqn:WKBeqn}) at a resonance to the surrounding area in the complex $\Omega$-plane is $\mathcal O(10^{-9})$.

We only report on results that we cross checked with the continued fraction method. 
{We have calculated some parts of the spectrum using a shooting method, where the differential equations (\ref{eqn:diff.second.order}) are solved with the proper boundary conditons (\ref{bc_horizon}) and (\ref{bc_infinity}). Although we have been able to reproduce the general behaviour of the spectrum with the mass, the precision obtained with this approach was not as good as the precision obtained with the continued fraction method and the third order WKB. Hence in the following figures we will only display results from these last two methods.}

\section{Results}\label{S3}

\subsection{An analytical approximation for large angular quantum number and mass}\label{sec.large.kappa}

Before presenting the numerical results obtained by applying the methods we have described, let us comment on an analytical result that can be derived from the WKB approach.
In the limit of large mass and angular quantum number ($\eta \gg 1$ and $|\kappa| \gg 1$), the second order equation (\ref{eq_WKB}) for $\nu_{1,2}$ becomes
\begin{eqnarray}
\frac{\mathrm d^2}{\mathrm d z^2} \nu_{1,2} + \left( \Omega^2 - \frac{\kappa^2 + \eta^2 x^2}{x^2} \, f \right) \nu_{1,2} = 0 \, . \label{eq_large_eta_kappa}
\end{eqnarray}
We can define an
effective potential 
\begin{eqnarray}
V_\text{eff}(x) = \frac{\kappa^2 + \eta^2 x^2}{x^2} f(x) = \left( \eta^2 + \frac{\kappa^2}{x^2} \right) f(x) \, . \label{eqn:eff.potential}
\end{eqnarray}
Note that this approximation to the effective potential is quadratic in $\kappa$ and thus unaffected by the sign of $\kappa$. This means that for large values of the angular number $l$, the two branches of modes that result from equation (\ref{kappa_def}) will become very close to each other.

The effective potential possesses an extremum determined by the equation
\begin{eqnarray}
\eta^2 (d-3) x^2 + \kappa^2 \left( d-1 - 2 x^{d-3} \right) = 0 \, . \label{eqn:large.mass.angular.extremum}
\end{eqnarray}
Hence following the standard procedure we can obtain a first order WKB approximation for the eigenvalue $\Omega$ in terms of the extremum of the potential, given by
\begin{eqnarray}
\Omega^2 = V_0 - \mathrm{i} \left( n + \frac{1}{2} \right) (-2 V_0^{(2)} )^{1/2} \, , \label{eqn:large.mass.angular.omega}
\end{eqnarray}
where $n$ is the overtone number.
This formula allow us to obtain an analytical approximation of $\Omega$ for large values of $\eta$ and $|\kappa|$. However, the relation has some limitations. For instance, when $\eta \gg 1$ and $|\kappa| \ll \eta$, the potential of equation (\ref{eq_large_eta_kappa}) does not have an extremum for finite $x>1$. Thus the above approximation (\ref{eqn:large.mass.angular.omega}) breaks down when $|\kappa|$ is small and $\eta$ is large.

However, for $\eta = 0$ the equation (\ref{eqn:large.mass.angular.omega}) becomes the eikonal $|\kappa| \gg 1$ limit of the massless Dirac modes, described before in \cite{Cho:2007zi}. Even more, in practice equation (\ref{eqn:large.mass.angular.omega}) is 
a reasonable approximation to the frequency $\Omega$ for all $\eta$ as long as $|\kappa| \gg 1$ and the extremum (\ref{eqn:large.mass.angular.extremum}) is not found at infinity. Note that because of the particular dependence of equation (\ref{eqn:large.mass.angular.extremum}) on the dimension $d$, it can be solved analytically only for $4\leq d \leq 7$ and $d=9$. For $d=8$ the extremum has to be determined numerically. 

We will now argue that the effective potential using the first order WKB approximation indicates the existence of frequencies with arbitrary small imaginary part for finite mass values $\eta$ in $d=4$ and $5$. If the second derivative of the potential vanishes in equation (\ref{eqn:large.mass.angular.omega}) and $V_0 > 0$, then the resulting frequency will be real. 
These conditions can be cast in the form
\begin{eqnarray}
0 &= \frac{(d-5) x^d + (d-1) x^3}{x^{d+3}} \, , \nonumber \\
\eta^2 &= \kappa^2 \frac{6 x^d + (1-d) d x^3}{(3-d)(2-d) x^5} \, .
\label{zero_pot}
\end{eqnarray}
For $d=4$ these result in a finite $x_{00}^{d=4} = 3$ and $\eta_{00}^{d=4} = |\kappa| / \sqrt{3}$
,
{where $x_{00}^{}$ and $\eta_{00}^{}$ are the zeros of the equations (\ref{zero_pot}) for the given value of $d$}.
For $d \ge 5$ the first equation does not have a finite solution $x_{00} > 1$. Still $x_{00}^{d \ge 5} \rightarrow \infty$ is a formal solution to the first equation. This results in a finite $\eta_{00}^{d=5} = |\kappa|$ only for $d=5$. For $d \ge 6$ also $\eta_{00}^{d \ge 6} \rightarrow \infty$ for the formal solution $x_{00}^{d \ge 5} \rightarrow \infty$.

The value of the potential at these points is in four dimensions $V^{d = 4}_\text{eff.} (x^{d=4}_{00}, \eta^{d=4}_{00}) = 8 \kappa^2 / 27 > 0$ and in five dimensions $V^{d=5}_\text{eff.}(x^{d=5}_{00}, \eta^{d=5}_{00}) = \kappa^2 > 0$. Thus in four and five dimensions one can find frequencies with vanishing imaginary part at finite mass $\eta$ in the first order WKB approximation. Of course in five dimensions the WKB method must break down close to these frequencies due to $x_0 \rightarrow \infty$ there. Although in the four dimensional case we do not see indications of a breakdown of the WKB method close to $x^{d=4}_{00} < \infty$, we will see that in practice both numerical schemes become troublesome close to $\eta^{d=4}_{00} = |\kappa| / \sqrt{3}$. However, as one approaches these points, in principle one should be able to generate frequencies with imaginary parts as small as desired for finite mass values. Note that the above behaviour is independent of the overtone number $n$ in the first order WKB approximation.

 Observe also that one can construct the effective potential given in equation (\ref{eqn:eff.potential}) from the geodesic equations of a massive particle of mass $m$. The Lagrangian $\mathcal{L}$ for a massive particle in the Schwarzschild-Tangherlini spacetime is given by
\begin{eqnarray}
\mathcal L = 1 = f(r) \dot{t}^2 - \frac{1}{f(r)} \dot{r}^2 - r^2 \dot{\Omega}^2 \, ,
\end{eqnarray}
where $\dot \Omega$ encodes all the angular dependencies. Of course one can choose a plane for the orbit and one has the constants of motion $l = r^2 \dot \Omega$ and $e = f(x) \dot{t}$. Substituting these in the above Lagrangian, multiplying with $m^2 \mu^2$, changing to the normalised quantities $\eta = m \mu, L = m l, x = r / \mu$ and rearranging gives
\begin{eqnarray}
\mu^2 \eta^2 \dot{x}^2 + \left(\eta^2 + \frac{L^2}{x^2} \right) f(x) = \eta^2 e^2 \, .
\end{eqnarray}
We can thus define the effective potential of geodesic motion
\begin{eqnarray}
V_\text{eff.}^\text{geo.} (x) =  \left(\eta^2 + \frac{L^2}{x^2} \right) f(x) \stackrel{L \leftrightarrow \kappa}{=} V_\text{eff.} (x) \, .
\end{eqnarray}
This should be expected, considering that to lowest order in $\hbar$ a WKB approximation for the Dirac equation results in the equations for geodesic motion \cite{Audretsch:1981wf, Rudiger:1981uu}.

\subsection{The case with large angular quantum number: numerical results for $l=10$}

In this section we present results for $l=10$, and different values of the spacetime dimension and fermionic mass. This can be seen as an example of the behaviour of the quasinormal modes for large values of the angular quantum number, {since as we will see below, the previous analytical result already describes very accurately the spectrum of modes. 
}

\begin{figure}
\includegraphics[width = 8.0cm]{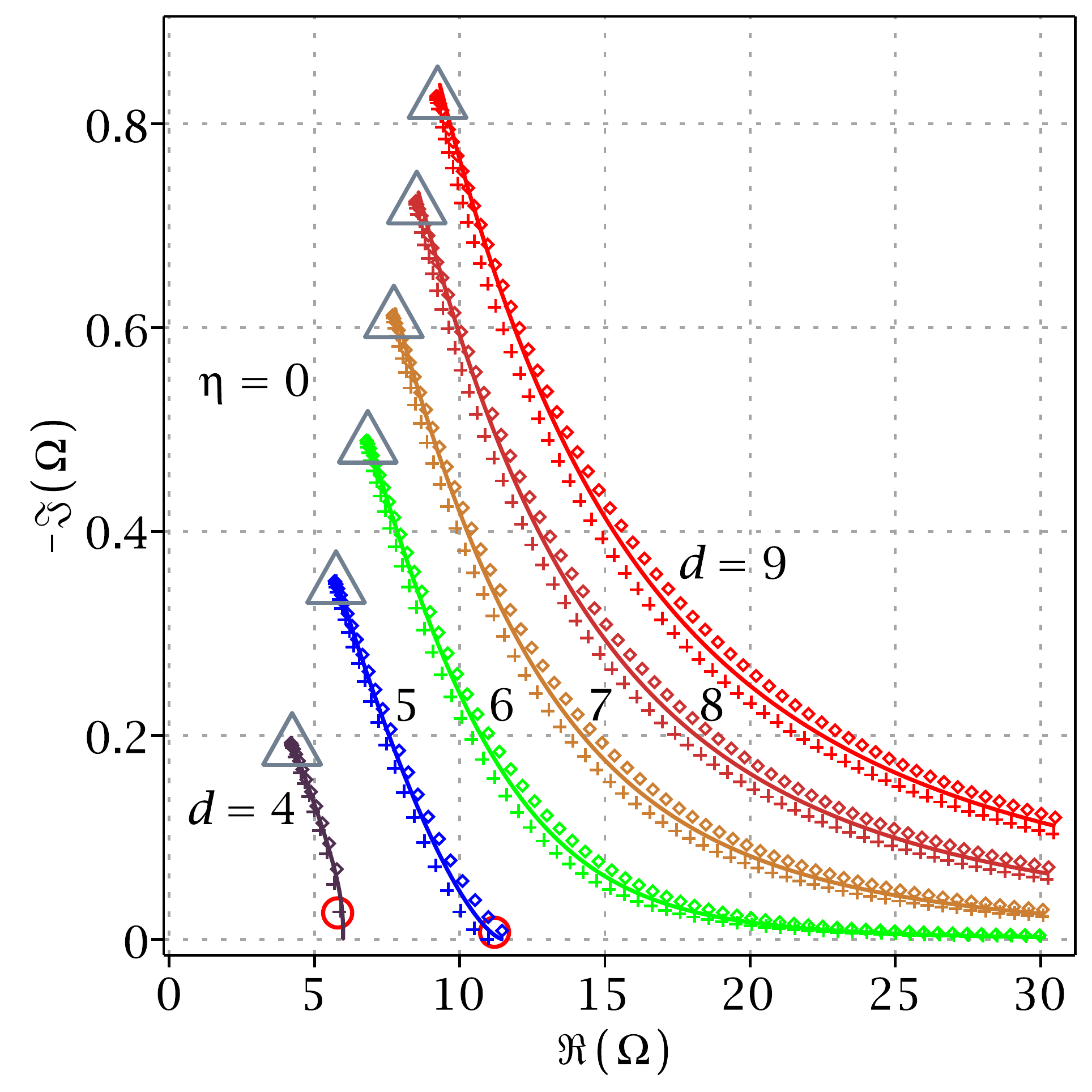}
\caption{\label{fig:RIOl10} 
The fundamental quasinormal modes 
in the complex $\Omega$-plane for $l=10$. The crosses and dots are the 
results from
the 3rd order WKB method with $\mathrm{sgn} (\kappa) = +1$ and $\mathrm{sgn} (\kappa ) = -1$, respectively. The difference in mass value between two adjacent points is $\Delta \eta = 0.5$. The solid lines are the analytical results from the large $\kappa$ approximation using the first order WKB approximation. 
From left to right in different colors $d=4, 5, 6, 7, 8, 9$. Marked with grey triangles are the quasinormal modes for $\eta = 0$. With increasing $\eta$, the value of the $|\Im(\Omega)|$ becomes smaller and the real part becomes larger. Around the red circles, for $d=4, 5$ the imaginary part becomes very close to zero, and the numerical methods break down.}
\end{figure}

The calculated quasinormal modes in the complex plane can be seen in Figure \ref{fig:RIOl10}, where we show the imaginary part versus the real part of $\Omega$. All these modes correspond to the fundamental state, for different values of the dimension $d$ and mass $\eta$. In different colors we show different spacetime dimensions $d$, ($d=4$ in purple, $d=5$ in blue, $d=6$ in green, $d=7$ in orange, $d=8$ in brown and $d=9$ in red). Each point represents a mode with a different value of the fermionic mass $\eta$, with the case $\eta=0$ (marked with a triangle) corresponding to the value with lowest frequency $\Re(\Omega)$, and largest $|\Im(\Omega)|$. Note that all these modes have $\Im(\Omega)<0$, so the Dirac field perturbation decays with time.
The crosses and points represent numerical values calculated using the third order WKB method with $\mathrm{sgn} (\kappa) = +1$ and $\mathrm{sgn} (\kappa ) = -1$, respectively. The solid lines represent the analytical results obtained from the large $\kappa$ limit to the potential (\ref{eqn:eff.potential}) in the approximation (\ref{eqn:large.mass.angular.omega}). We can see that even though in the approximation we assumed Dirac fields with $\eta \gg 1$, in practice the massless case is very well approximated, and the analytical approximation works very well for all values of the mass $\eta$: the relation (\ref{eqn:large.mass.angular.omega}) works as an eikonal approximation for the Dirac quasinormal modes, even for small and intermediate values of the mass. 

\begin{figure}
\includegraphics[width = 8.0cm]{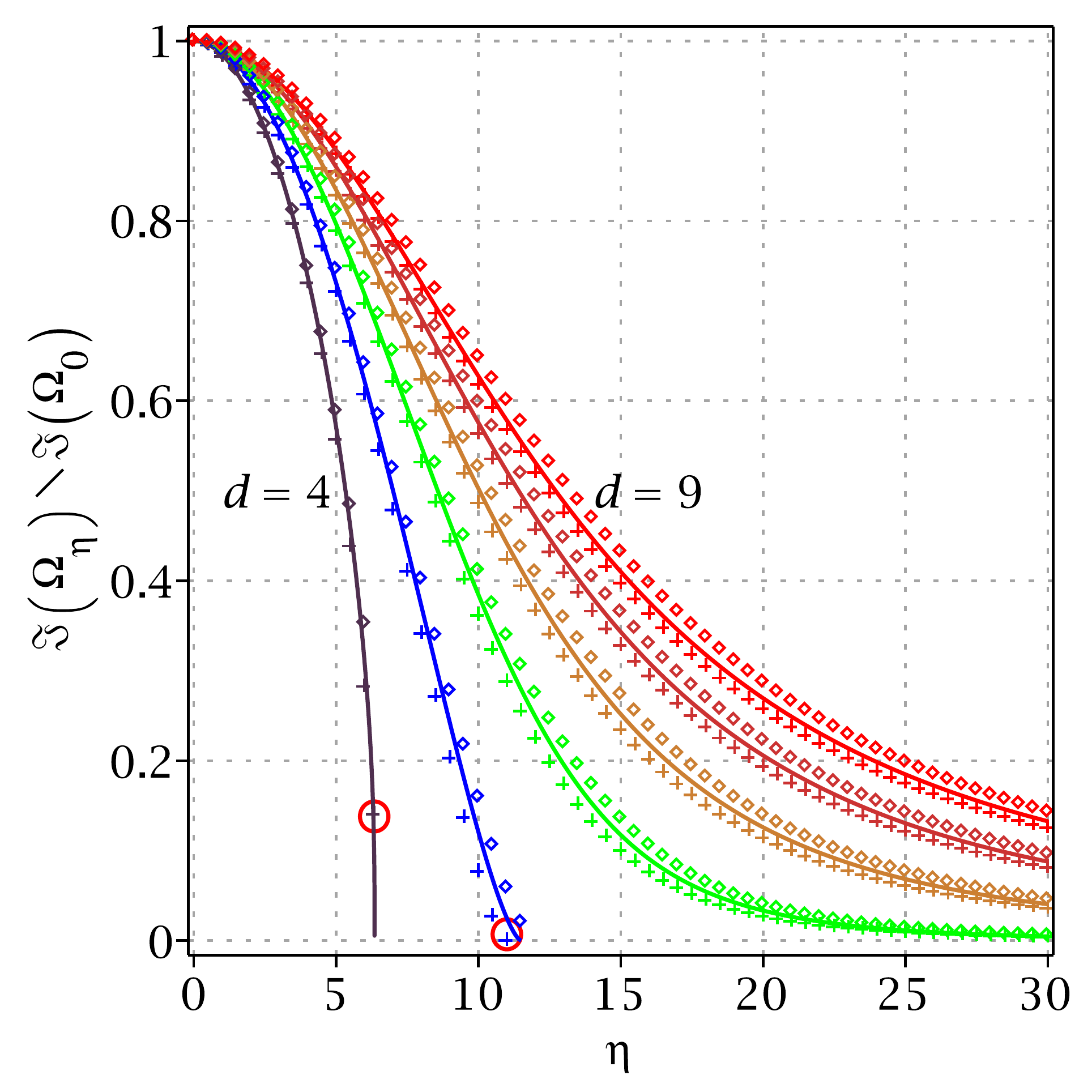}
\caption{\label{fig:Ibeh} 
Similar figure to the previous one, showing the
imaginary part of $\Omega$ over the mass parameter $\eta$ for the fundamental $l=10$ mode. We normalize with respect to the imaginary part of the frequency for $\eta=0$.  
}
\end{figure}

In Figure \ref{fig:RIOl10} we can see that increasing the mass of the Dirac field has the effect of increasing the frequency of the mode $\Re(\Omega)$, while the absolute value of the imaginary part $\Im(\Omega)$ decreases (meaning that the damping time of the perturbation increases). Also we can appreciate how increasing the dimension has the generic effect of increasing the value of $\Re(\Omega)$ and $|\Im(\Omega)|$. With regards to the sign of $\kappa$, in this figure one can observe that for a fixed value of the mass $\eta > 0$
the analytical approximation lies always inbetween the full numerical values obtained for each one of the branches. 
This means that
$\Re (\Omega_{\mathrm{sgn}(\kappa)=+1}) < \Re (\Omega_\text{ana}) < \Re (\Omega_{\mathrm{sgn}(\kappa)=-1})$ and $|\Im (\Omega_{\mathrm{sgn}(\kappa)=+1})| < |\Im (\Omega_\text{ana})| < |\Im (\Omega_{\mathrm{sgn}(\kappa)=-1})|$
, where $\Omega_\text{ana}$ are the quasinormal modes from the large $\kappa$ approximation using first order WKB.
In any case, as it was expected in the previous section, for $l=10$ both signs of $\kappa$ are always close to the analytical approximation.

Note that the cases with $d=4$ and $d=5$ possess a different behaviour from the rest of dimensions considered. In $d=4, 5$, the numerical analysis indicates that there exists a value of $\Re(\Omega)$ for which the $\Im(\Omega)$ reaches a critical value, as was indicated by the argument made in section \ref{sec.large.kappa}. However, the numerical methods break down in this region and we mark these values with a red circle in Figure \ref{fig:RIOl10}. This is not the case in $d\geq6$, where $|\Im(\Omega)|$ decreases smoothly as the frequency increases.

In Figure \ref{fig:Ibeh} we present the scaled value of $\Im(\Omega)$ (normalized to the massless value of the imaginary part $\Im(\Omega_0)$). Here we can appreciate the critical behaviour for $d=4,5$. In $d=4$ for a critical value of the fermionic mass $\eta_c$, the imaginary part of $\Omega$ goes to zero in the analytical approximation. The numerical results show that the branch of modes disappears at a finite $\eta$, which is slightly smaller than the critical fermionic mass predicted by the analytical appriximation. 
In $d=5$ however, the numerical analysis concides very well with the analytical approximation, showing that the $\Im(\Omega)$ can become arbitrarily small as we reach the critical value of the fermionic mass. However the numerical analysis becomes problematic very close to the critical value, indicating that this branch of modes also disappears before reaching this value.

{We want to comment here that the existence of this critical value of the fermionic mass in the spectrum of the $d=4$ case was already noted in \cite{Cho:2003qe} by analysing the shape of the potential. Also it was observed that more massive fields lead to more slowly damped modes. Our results are compatible with these observations.}

In the cases with $d\geq6$ the behaviour changes, and we can see that the modes exist for arbitrary values of the fermionic mass. The imaginary part decreases monotonically to zero as the mass becomes larger and larger. Hence the damping time of these modes can be increased without limit for very large values of the mass $\eta$. 

\begin{figure}
\includegraphics[width = 8.0cm]{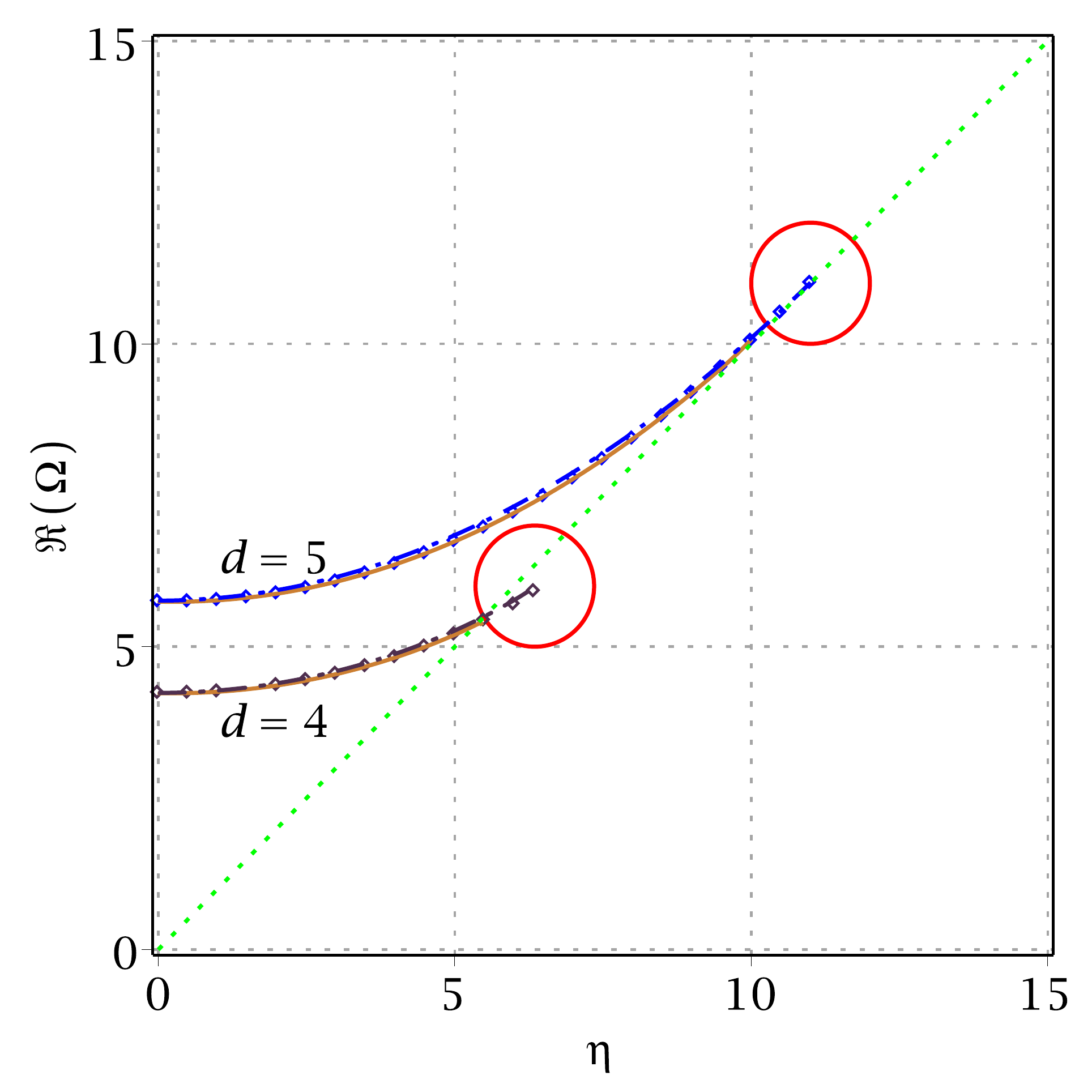}
\caption{\label{fig:d45_Phase} Real part of the fundamental modes for $d=4, 5$ and $l=10$ ($\mathrm{sgn}(\kappa) = +1$), as a function of the mass $\eta$. The dotted green line marks $\Re (\Omega) = \eta$. The dash-dotted curves are the analytical approximation in the large $\kappa$ limit using the first order WKB approximation. The circle points are the results from the 3rd order WKB method and the solid orange curves are the results from the continued fraction method. 
The red circles mark the region where the WKB, continued fraction method and analytical approximation break down. 
}
\end{figure}

Some insight into the particular behaviour of the cases with $d = 4, 5$ can be obtained by studying the dependence of $\Re(\Omega)$ on the mass $\eta$. In Figure \ref{fig:d45_Phase} we show the real part of $\Omega$ as a function of $\eta$ for these two values of the spacetime dimension for positive $\kappa$. Here we show the modes computed using both the WKB method (circle points) and the continued fraction method (solid line). The dashed lines represent the analytical approximation (\ref{eqn:eff.potential}) and (\ref{eqn:large.mass.angular.omega}), and the green line marks the limit in which $\Re(\Omega)=\eta$. 

The numerical methods break down precisely when this line is approached (red circles), and we cannot trust results obtained beyond this point. Although the analytical approximation can be extended to generate some modes beyond $\Re(\Omega)=\eta$, in this case it is probably an artifact of the approximation. 

\begin{figure}
\includegraphics[width = 8.0cm]{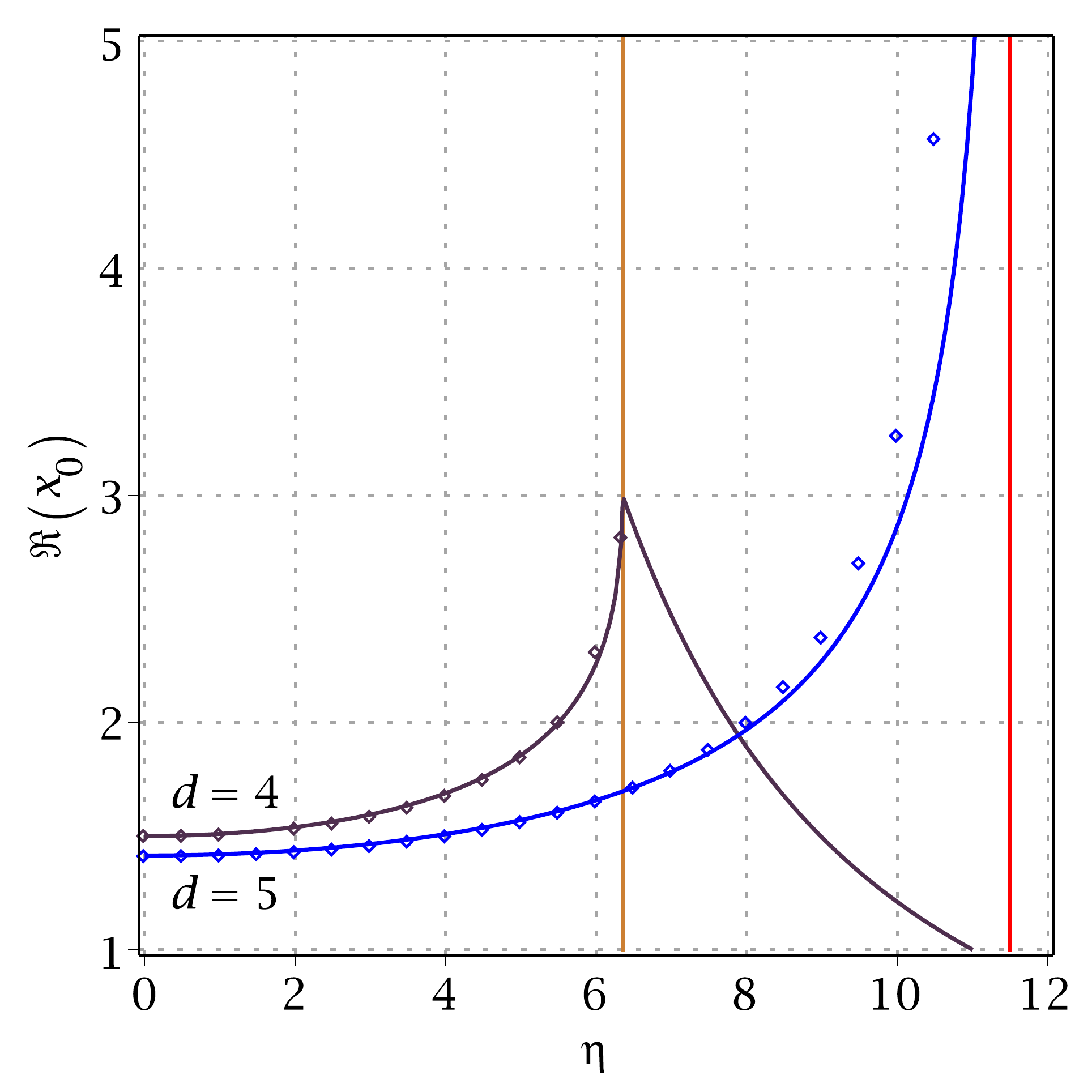}
\caption{\label{fig:d45_Rex} Real part of the point $x_0$ for which $V_{\text{eff}}^\prime (x_0) = 0$ over the mass $\eta$ for $l=10$ ($\mathrm{sgn}(\kappa) = +1$). The black and blue solid curves are the large $\kappa$ approximation for $d=4, 5$ respectively. The circle points were obtained using the full effective potential. The vertical solid red line marks the values of the mass for which $x_0$ diverges in five dimensions at $\eta_{00}^{d=5} = \kappa$. The orange vertical line marks the value of $\eta_{00}^{d=4} = \kappa / \sqrt{3}$.
}
\end{figure}

\begin{figure}
\includegraphics[width = 8.0cm]{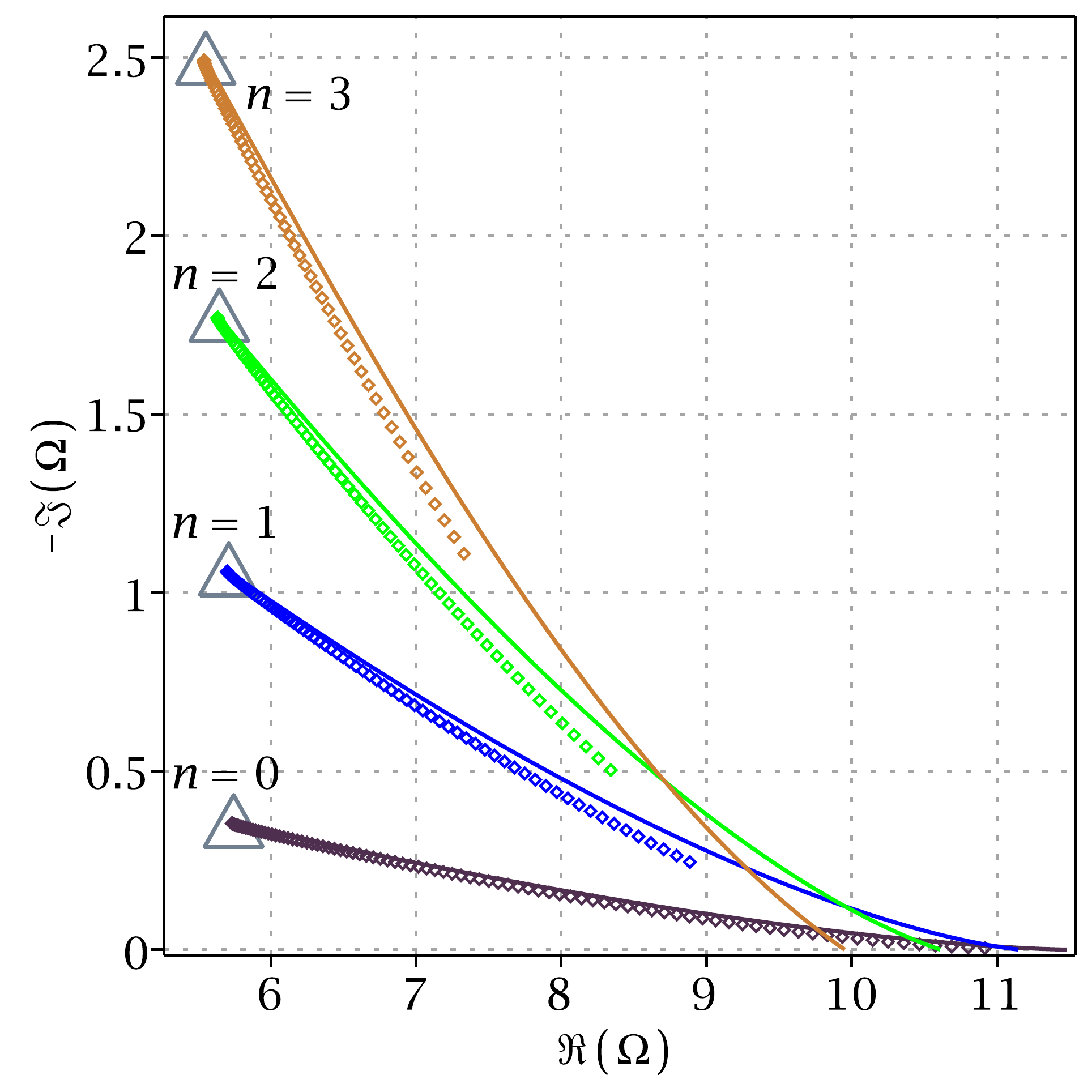}
\caption{\label{fig:d5nRIO} Imaginary part of $\Omega$ over the real part of $\Omega$ for different overtone numbers $n$ for the five dimensional case with $l = 10$ and $\mathrm{sgn}(\kappa) = +1$. From bottom to top $n=0,1,2$ and $3$. The circles are the numerical results using the third order WKB approximation using the full effective potential. The solid lines are the analytical values in the large $\kappa$ approximation of the effective potential also using the third order WKB approximation. For the full potential the numerical values are displayed up to the point the WKB and continued fraction method begin to diverge. Marked with grey triangles are the points with $\eta = 0$. 
}
\end{figure}

\begin{figure}
\includegraphics[width = 8.0cm]{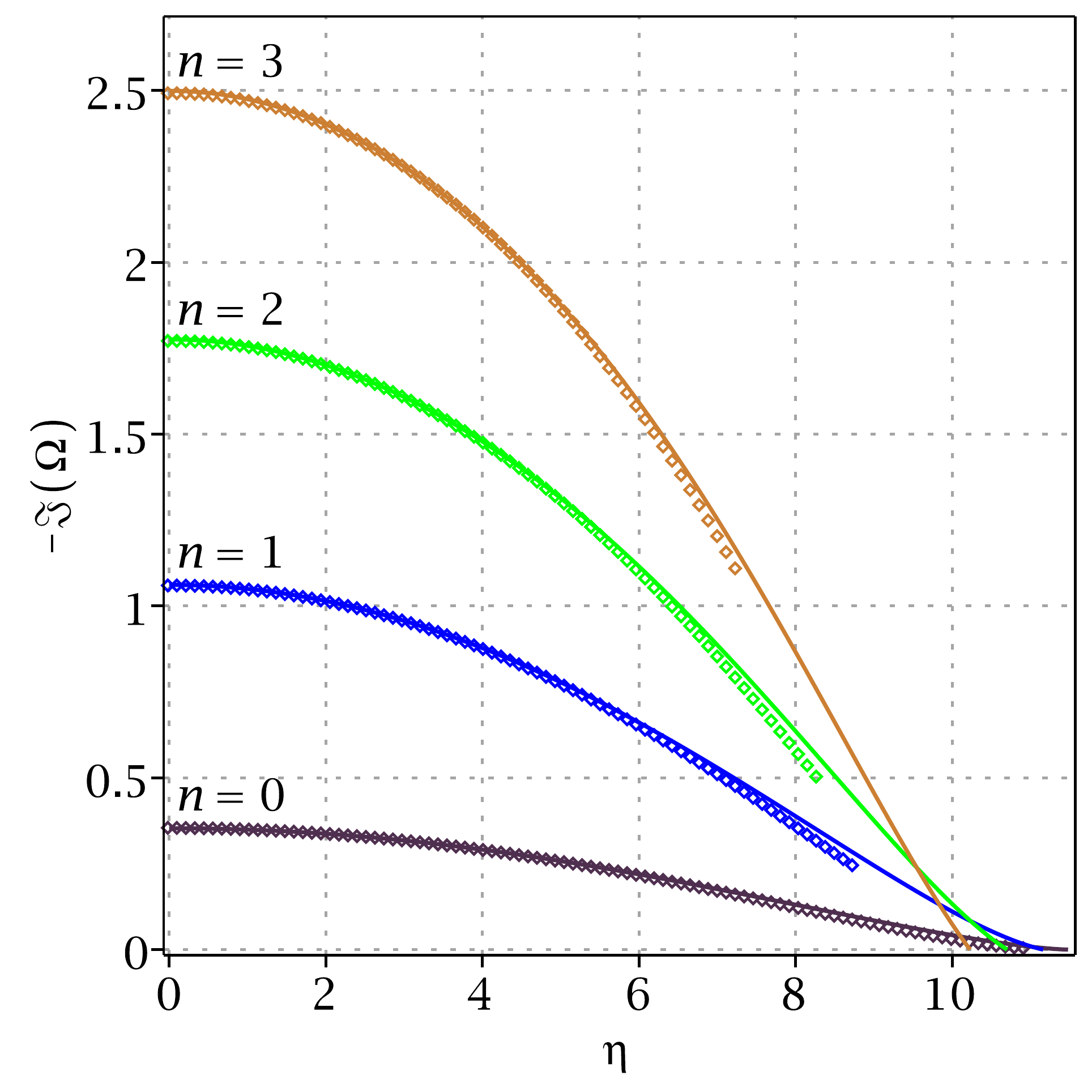}
\caption{\label{fig:d5nbeh} Similar figure to the previous one, here presenting the imaginary part of $\Omega$ over the mass $\eta$ for different overtone numbers $n$ for the five dimensional case with $l = 10$ and $\mathrm{sgn}(\kappa) = +1$. 
}
\end{figure}

In order to show this singular behaviour, in Figure \ref{fig:d45_Rex} we present the real part of the point $x_0$ satisfying $V_{\text{eff}}^\prime (x_0) = 0$, as a function of $\eta$ for $\mathrm{sgn}(\kappa)=+1$. The black ($d=4$) and blue ($d=5$) lines correspond to the analytical approximation (\ref{eqn:eff.potential}), while the circle points correspond to values calculated using the complete effective potential from equation (\ref{eqn:completeeffpot}). We can see in this Figure that the real part of $x_0$ diverges for $d=5$ as we approach the critical value of the mass $\eta_{00}^{d=5} = \kappa$ (marked with vertical red line). One can also see that although the values for $\Re(x_0)$ stay finite in $d=4$, there is a transition to another branch of zeros of the potential. However the numerical methods cannot obtain results with good enough precision once this second branch of zeros is reached.

Hence, since
both numerical methods cannot generate modes with good precision close to $\Re(\Omega)=\eta$, and the analytical approximation breaks down in $d=5$ when this limit is crossed,
we conjecture 
that this branch of fundamental $l=10$ modes ceases to exist in $d=4, 5$ when the mass of the Dirac field is large enough and the $\Re(\Omega)=\eta$ limit is reached.

In a summary, from the results we have presented in this section we can learn the following:

For $d\ge6$, as the mass of the fermionic field is increased, the imaginary part of the $l=10$ fundamental mode decreases. The mode seems to exist for arbitrary large values of the mass $\eta$, and $\Im(\Omega)$ can become arbitrarily small. Hence it is possible to find massive modes with arbitrary large damping times (which we can interpret as quasistationary states for large fermionic mass). The analytical approximation of the spectrum gives very reasonable results for all values of the mass. Since in this regime the frequency increases linearly with the mass, the slowly damped field will oscillate very rapidly with $\Re(\Omega)\simeq \eta \gg 1$

For $d=4, 5$, the behaviour is different from the higher dimensional case, since the branch of modes originating from the massless fundamental mode seems to disappear at a finite value of the fermionic mass. However, both the approximation and the numerics indicate that this limit is singular, and the mode is lost before reaching the critical mass. 

In $d=4$, although the analytical approximation predicts that arbitrary small values of $\Im(\Omega)$ can be obtained at a finite critical $\eta$, the branch of modes in the numerical approach is lost before this point is reached. 

In $d=5$ surprisingly, the analytical approximation of the spectrum matches very well with the numerics for all values of the mass lower than the critical $\eta$.
For mass values arbitrarily close to this critical point, in principle very small values of $\Im(\Omega)$ can be generated. These can be interpreted as quasistationary states with intermediate values of the fermionic mass. In this regime the field oscillates with a frequency $\Re(\Omega)\simeq \eta$.  

{Note that for larger values of $l$, the full numerical results are in fact closer to the spectrum predicted by the analytical formula (for $l=10$ the numerics only deviate from the analytical formula less than a $5\%$, for all values of the mass and dimension considered). So the analytical aproximation gives very reasonable results for $l\ge 10$. 
}

As a final note in this section we would like to point that the results we have presented here coincide, within the expected precision, with the results obtained in the recent work \cite{KonoplyaKerr2017}. Here the authors study the massive Dirac modes in the background of the Kerr black hole, and obtain similar results for the spectrum of the Dirac field in the background of a Schwarzchild black hole in the static limit.

\subsection{The behavior for different overtone number $n$: numerical results for $l=10$ in five dimensions}

As we commented in the previous section \ref{sec.large.kappa}, in the lowest order approximation and independently of the overtone number, one can see that $\Omega \rightarrow \kappa + \mathrm{i} \cdot 0$ for $\eta \rightarrow \eta_{00}^{d=5} = |\kappa|$,  with $n<l$.
However it is not clear if this feature is retained for all overtone numbers when one goes beyond the analytical approximation. To investigate this issue, in this section 
we explore how the overtone number $n$  affects the critical behaviour of the $l=10$ quasinormal modes in $d=5$ for $\mathrm{sgn}(\kappa)=+1$.

In Figure \ref{fig:d5nRIO} we present the imaginary part vs. the real part of the frequency for different overtone numbers $n$. The circles are the numerical results using the third order WKB approximation  (\ref{eqn:WKBeqn}) using the full effective potential  (\ref{eqn:completeeffpot}), and cross-checked with the CF method. The solid lines are the analytical values in the large $\kappa$ approximation of the effective potential (\ref{eqn:eff.potential}) also using the third order WKB approximation (\ref{eqn:WKBeqn}). The analytical approximation is shown for the whole range of its validity. However, for $n\ge 1$, the branches of modes calculated from the numerics do not reach the $\Im{(\Omega)}=0$ value. 

In Figure \ref{fig:d5nbeh} we make a similar plot, showing the imaginary part of the frequency vs. the mass $\eta$ for different overtone numbers $n$. 
Again it can be seen that for $n\ge 1$, the branches of modes calculated from the numerics stop at a value of $\eta$ always below the critical value of $\eta$ predicted from the analytical approximation.

In the analytical approximation, for all displayed overtone numbers, the frequencies develop arbitrary small imaginary parts for finite mass $\eta$. For the analytical approximation we can thus define a critical mass $\eta_c(n)$ for which $\Im (\Omega) \rightarrow 0$. One can observe that for all displayed overtone numbers $\eta_c(n) < \eta_{00}^{d=5} = \kappa = 11.5$. 

However the numerical analysis using WKB and CF methods indicates that in fact the branch of quasinormal modes disappears before reaching this particular value of $\eta_c(n)$. Beyond this value of the fermionic mass, our current methods are not able to produce quasinormal modes with good enough precision. 

Thus the indicated universal behaviour for the frequency is not completely retained by higher order WKB methods. In fact, as we have seen, the critical mass value depends on the overtone number, $\eta_c(n)$, being smaller than the predicted value from the lower order WKB method, $\eta_{00}^{d=5}$. Even more, the full numerical approach indicates that the branch of modes disappears at even lower values of the fermionic mass 
\footnote{A similar behaviour with the overtone number $n$ seems to be present in the $d=4$ case. However the WKB and continued fraction methods don't allow us to obtain results with good enough precision, and another approach should be used to fully understand this case.}.

\subsection{Numerical results for $l=0$}

In this section we present the spectrum for angular quantum number $l=0$. We calculate the fundamental and first excited 
modes for $4 \le d \le 9$ with the continued fraction method.

\begin{figure}
\includegraphics[width = 8.0cm]{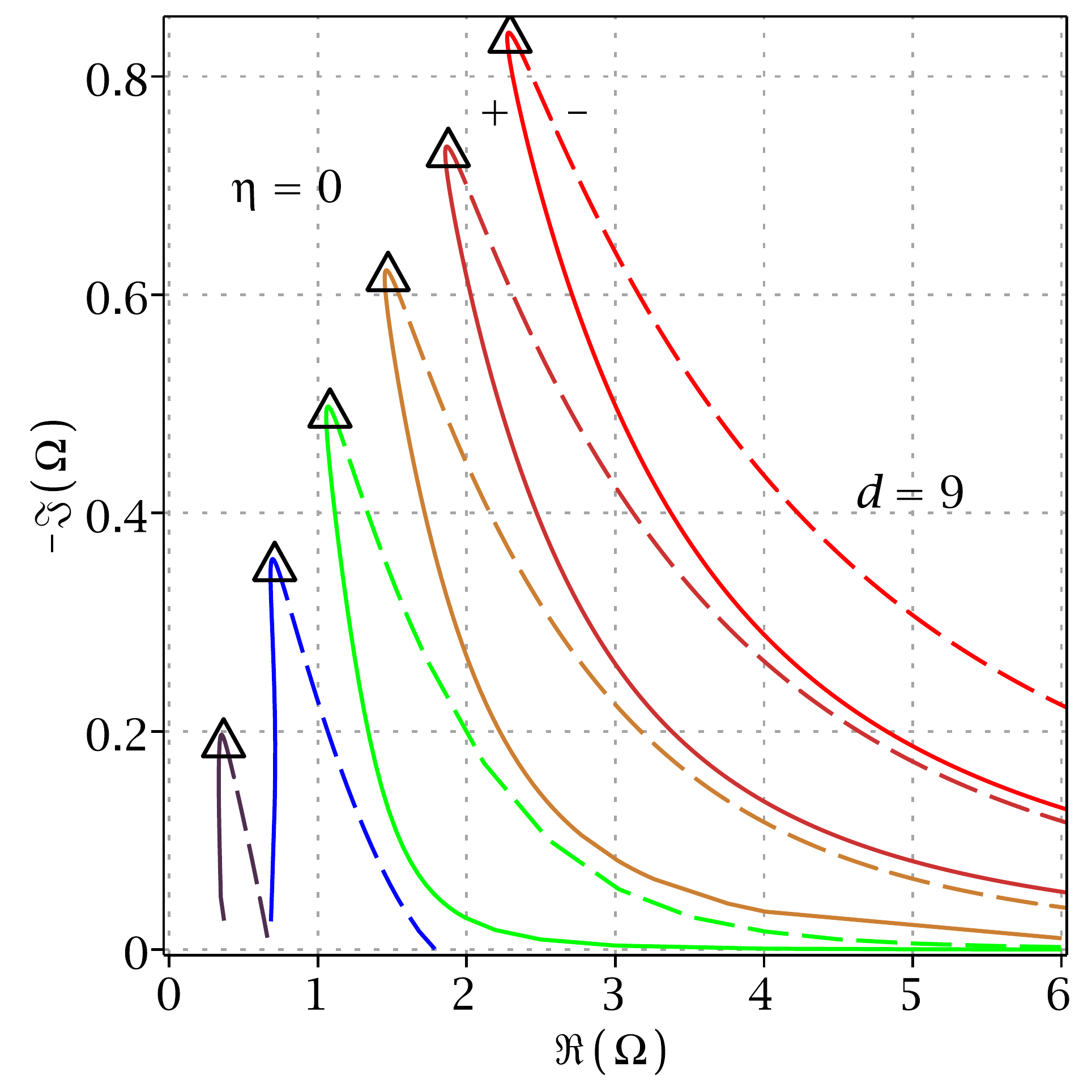}
\caption{\label{fig:RIOl0n0} Fundamental modes in the complex $\Omega$-plane for $l=0$, using the continued fraction method. From the left to the right we have $d=4,5,6,7,8$ and $9$. Shown in continuous lines are the values for $\mathrm{sgn}(\kappa)=+1$ and in dashed lines the values for $\mathrm{sgn}(\kappa)=-1$.
Marked with black triangles are the quasinormal modes for $\eta=0$. We only show modes with a relative error of $0.5\%$, except in $d=4$ and $5$ close to the real axis, where the  numerical method breaks down, and we relax the condition to a relative error of $2 \%$.
}
\end{figure}

\begin{figure}
\includegraphics[width = 8.0cm]{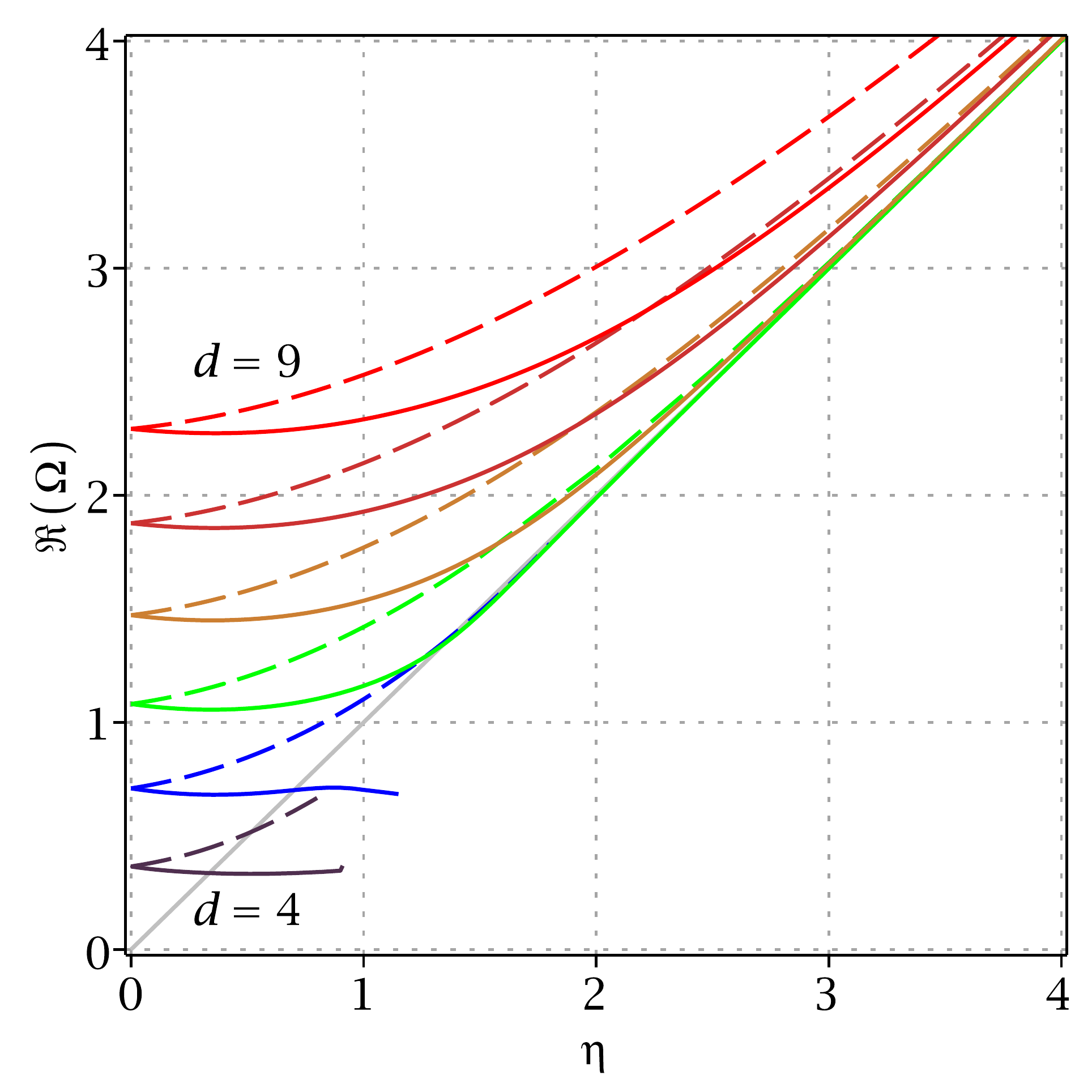}
\caption{\label{fig:Rbehl0n0} A similar figure to the previous one, showing the real part of the fundamental mode as a function of the mass for $l=0$, and different values of the spacetime dimension $d$. 
In grey we include the line $\Re(\Omega)=\eta$.}
\end{figure}

In Figure \ref{fig:RIOl0n0} we present $\Im(\Omega)$ vs $\Re(\Omega)$ for the fundamental mode. Note that all the modes calculated have negative imaginary part, meaning they also decay in time. 
Marked with black triangles are the quasinormal modes for $\eta=0$. 
Two branches of modes bifurcate from these points: one for $\mathrm{sgn}(\kappa)=+1$, shown with continuous lines, and one for $\mathrm{sgn}(\kappa)=-1$, shown with dashed lines. For fixed values of $\eta$, the branch with $\mathrm{sgn}(\kappa)=+1$ presents always smaller values of the real and imaginary parts of $\Omega$ than the branch with $\mathrm{sgn}(\kappa)=-1$. Also one can see that
as the dimension $d$ increased, the fundamental mode increases both in $\Re(\Omega)$ and in $|\Im(\Omega)|$, and approximately by the same amount for a fixed value of the mass $\eta$.

When $\eta$ is increased, the absolute value of the imaginary part of the frequency $|\Im (\Omega)|$ becomes generally smaller. 
For $6 < d \le 9$, this seems to lead to modes that approach the real axis asymptotically, independently of the sign of $\kappa$. 
However,
in four and five dimensions
our results indicate that the modes stop existing at a certain value of $\eta$, right before hitting the real axis.
These features are qualitatively very similar to the behaviour obtained in the large $l$ limit, and in particular for $l=10$ in the previous section (see Figure \ref{fig:RIOl10}) and to the generic features of the ground state of a scalar field \cite{Zhidenko:2006rs}. Interestingly, in Figure \ref{fig:RIOl0n0} we can observe that 
the branch of modes in $d$ dimensions with negative $\kappa$ seems to approach the branch of modes in $d+1$ dimensions with positive $\kappa$. 

Here again the numerical methods we employ break down for very small values of $|\Im (\Omega)|$, and in practice we cannot generate arbitrary small values of the imaginary part (specially in the positive $\kappa$ branches).
As the mass approaches the critical value, the results produced from the equation for $\psi_1$ in relation (\ref{eqn:CF}) deviates more and more from the results for $\psi_2$, and the depth of the continuous fraction method has to be increased in order stabilize the computed frequencies. We only show results that we are able to cross-check using both equations, with less than a $0.5\%$ of difference in the calculated $\Omega$ (except close to the critical points of the $d=4, 5$ cases with positive $\kappa$, where we relax it to a $2\%$ of difference).

In Figure \ref{fig:Rbehl0n0} we show the real part of the frequency as a function of the mass.
In this figure we can clearly see that for large $\eta$, the branch of negative $\kappa$ of a given dimension $d$ approaches the branch of positive $\kappa$ in $d+1$ dimensions. This is thus also true for the imaginary part.
Another feature we can observe in this figure is that, in the $\mathrm{sgn}(\kappa)=+1$ branch, the minimum value of the frequency no longer resides at the massless case, but at some configuration with non-zero fermionic mass.
Interestingly, for $d=4, 5$, and for $d=6$ with $\mathrm{sgn}(\kappa)=+1$, the real part of the frequency can become smaller than the mass of the field, crossing the line $\Re(\Omega)=\eta$. This is different from what we observed for $l=10$ in Figure \ref{fig:d45_Phase}. Thus in some cases the Dirac field can be trapped in the gravitational field of the black hole.

\begin{figure}
\includegraphics[width = 8.0cm]{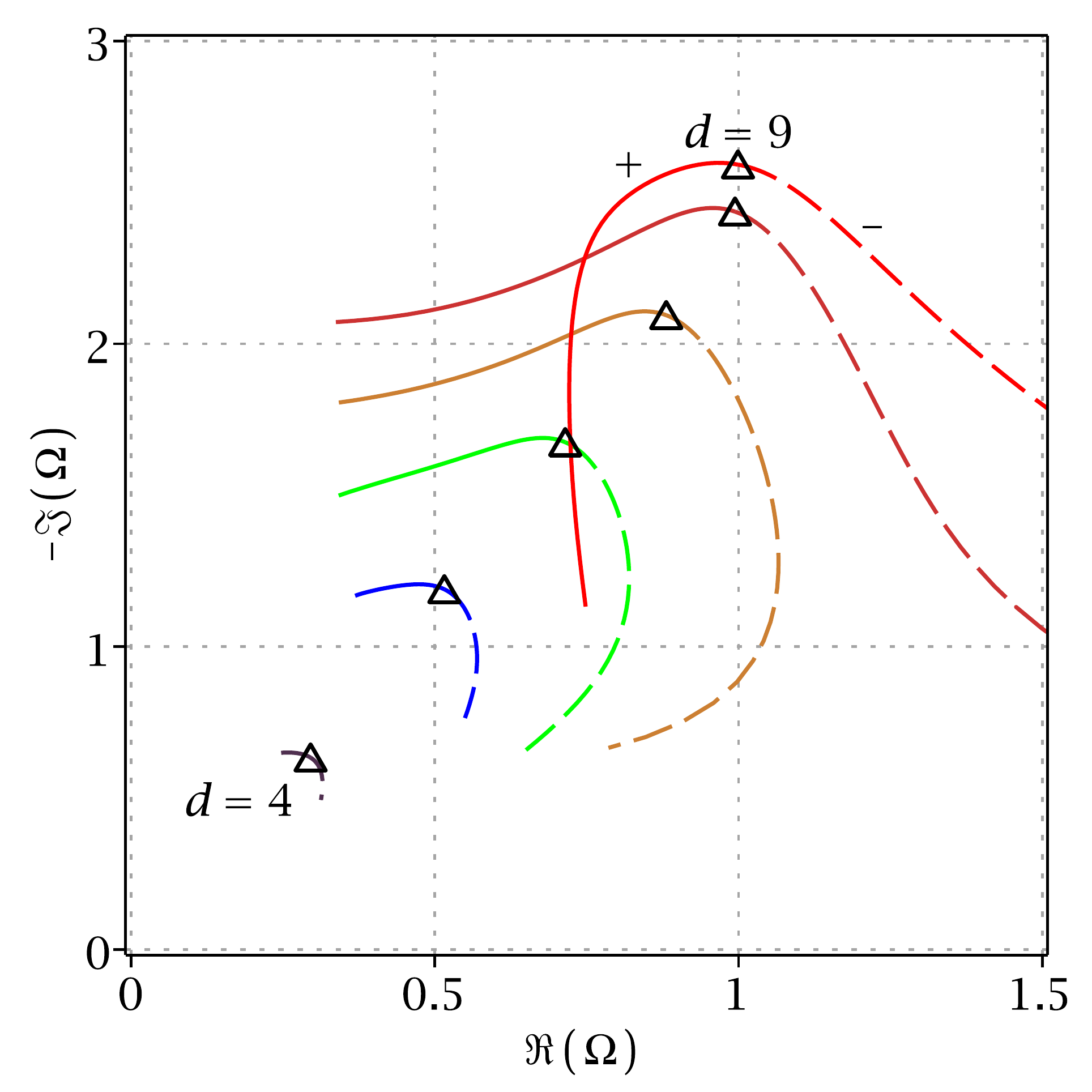}
\caption{\label{fig:RIOl0n1}  First excited quasinormal modes in the complex $\Omega$-plane for $l=0$, using the continued fraction method. From the bottom left to the upper right $d=4,5,6,7,8$ and $9$. Shown in continuous lines are the values for $\mathrm{sgn}(\kappa)=+1$ and in dashed lines the values for $\mathrm{sgn}(\kappa)=-1$.
Marked with black triangles are the quasinormal modes for $\eta=0$. 
We show modes with relative error smaller than $0.5\%$.
}
\end{figure}

\begin{figure}
\includegraphics[width = 8.0cm]{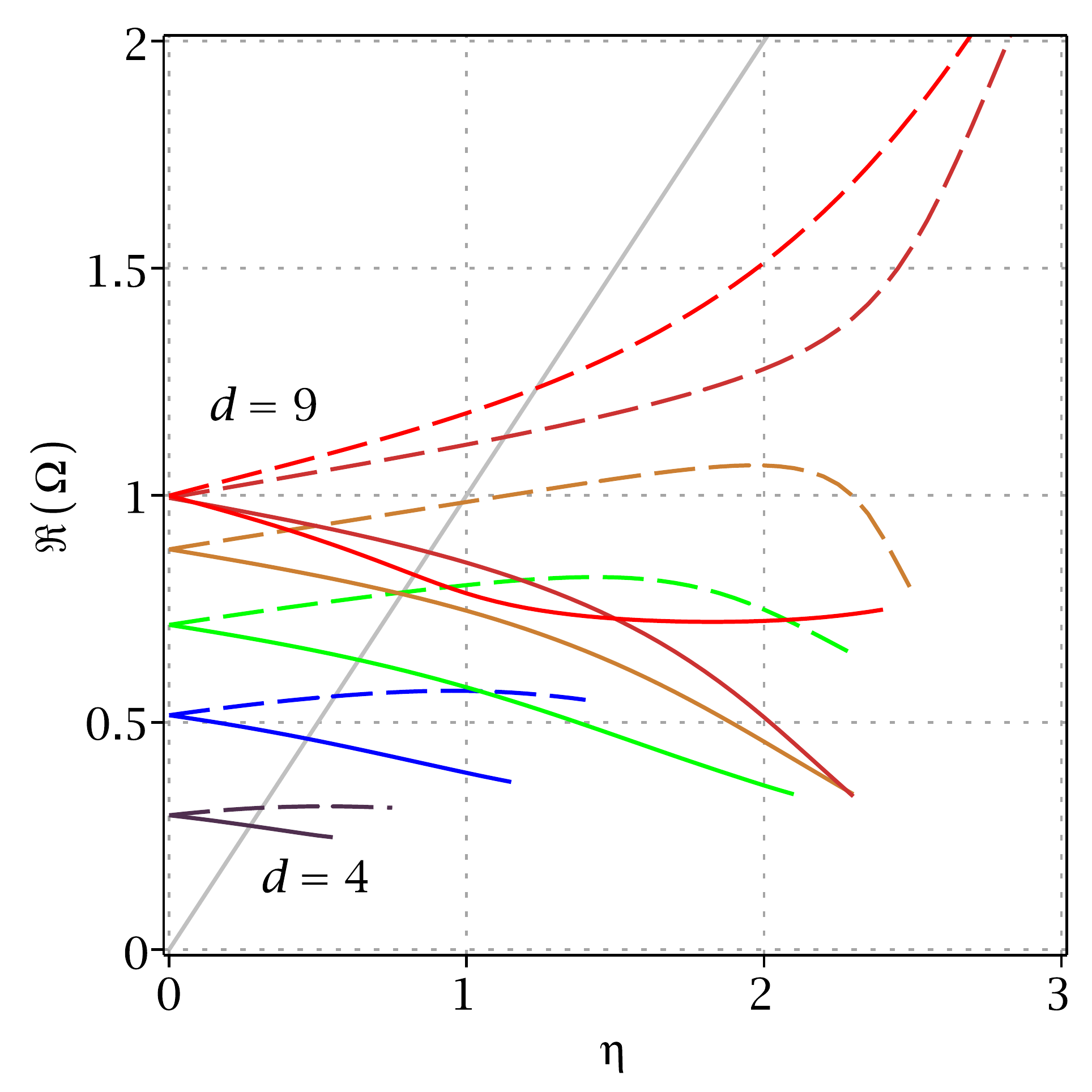}
\caption{\label{fig:Rbehl0n1} Similar figure to the previous one, showing the real part of the quasinormal modes for the  first excited modes for $l=0$. 
In grey we mark the line $\Re(\Omega)=\eta$.}
\end{figure}

As in the previous section, we would like to point out here that the results we have presented for the $d=4$ case coincide with the results obtained in the recent work \cite{KonoplyaKerr2017} within the required precision.

To conclude this section, we present some results regarding the first excitation ($n=1$) of the $l=0$ modes, although for large values of the mass the continued fraction method does not allow us to obtain the same precision as with the fundamental mode. In Figure \ref{fig:RIOl0n1} we show $\Im(\Omega)$ vs $\Re(\Omega)$ for the first excited mode, and in Figure \ref{fig:Rbehl0n1} we show the real part of $\Omega$ as a function of the mass for these modes. It is worth noting that some of the branches of the first excited mode of the spinor field possesses the generic behaviour of a vector field mode, see for example \cite{Konoplya:2005hr}. In the positive $\kappa$ branches of the $d=4...8$, increasing $\eta$ decreases the real part of the frequency $\Re( \Omega)$, while the imaginary part does not change much. 
In the negative $\kappa$ branches however, the real part increases, and only starts to decrease for relatively large values of the mass in the $d=4..7$ cases. Interestingly, the $d=8$ branch with negative $\kappa$, and both branches with $d=9$ seems to avoid the imaginary axis, so the similarity with the vector field modes is lost in these particular cases.

\section{Conclusion}\label{S4}

In this paper
we have calculated the quasinormal modes of the massive Dirac field in the Schwarzschild-Tangherlini black hole in $d=4$ to $9$. We have implemented the calculation of the modes with two independent methods: the continued fraction method, and the third order WKB method. In addition, we have obtained an analytical approximation of the spectrum, which formally applies to the case of large fermionic mass and large angular quantum number. However in practice, we have seen that the approximation works rather well for arbitrary values of the mass, provided the angular quantum number is large enough. 

As an example, we have investigated carefully the particular case of the $l=10$ fundamental mode. The frequency and the damping time of the mode increases to arbitrarily large values as the mass of the Dirac field increases. This is a feature that has been observed in other massive fields. Interestingly, for $d=4, 5$, the mode seems to disappear 
as the damping time of the mode rises at a critical value of the mass with $\eta=\Re(\Omega)$. While for $d=4$, the mode seems to disappear at a finite value of $\Im(\Omega)$, while for $d=5$ the damping time seems to grow arbitrarily as the critical value of the fermionic mass is reached. For $d\ge6$ however, the mode seems to exist for arbitrary values of the mass. These results, together with the analytical approximation obtained from the limit of large mass and angular quantum number, indicate that quasistationary perturbations for intermediate values of the mass and frequencies can be found for $d=4, 5$ for large values of the angular quantum number, while quasistationary perturbations with very large values of the mass can be found for $d\ge6$ with also large values of the frequency.

In addition, the effect of the overtone number has been explored for the particular case of $d=5$ and $l=10$. Interestingly, the full numerical analysis deviates significantly from the eikonal approximation as the mass is increased, and indicates that the branches of excited modes reach a critical value of the fermionic mass where they cease to exist. This critical value of the mass decreases with the overtone number.

We also present results for $l=0$ and the first two overtone numbers, $n=0, 1$. The behaviour of the Dirac field was analogous to a scalar field for the fundamental mode, and similar to a vector field for the first excitation for $d < 8$ and $d = 8$ with positive $\kappa$.
Also there exist gravitationally trapped modes with the real part of the frequency smaller than the mass $\Re(\Omega) < \eta$.

We were also able to observe, that the spectrum for the massive spinor depended on the sign of $\kappa$. In general the inequality $\Re (\Omega_{\mathrm{sgn}(\kappa)=+1}) < \Re (\Omega_{\mathrm{sgn}(\kappa)=-1})$ for a fixed value of mass $\eta > 0$ seems to hold.

The disappearance of the modes when the mass reaches a critical value may indicate the starting of another branch of modes for higher values of the mass. This branch of modes could have very small values of the imaginary part. However, another numerical approach is probably necessary, since with our current methods we are not able to study such values of the eigenmodes with large values of the fermionic mass. For instance, it may be more appropriate to change to another representation of the Dirac spinor.

\section{Acknowledgements}
The authors want to thank Jutta Kunz, Roman Konoplya and Alexander Zhidenko for discussions.
CK and JLBS gratefully acknowledge support by the DFG funded Research Training Group 1620 ``Models of Gravity'', and the FP7, Marie Curie Actions, People, International Research Staff Exchange Scheme (IRSES-606096). 

\bibliography{QNMDiracTang.bbl}


\end{document}